\documentclass[twocolumn]{aastex701}

\newcommand{\code}[1]{\texttt{#1}}

\begin{document}

\title{Isotopic Ratios in the Disk of HD 163296}

\author[0000-0001-8642-1786,sname='Qi']{Chunhua Qi}
\affiliation{Institute for Astrophysical Research, Boston University, 725 Commonwealth Avenue, Boston, MA 02215, USA}
\email[show]{cqi1@bu.edu}

\author[0000-0003-1526-7587,sname='Wilner']{David J. Wilner} 
\affiliation{Center for Astrophysics \textbar\ Harvard \& Smithsonian, Cambridge, MA 02138, USA}
\email{dwilner@cfa.harvard.edu}

\author[0000-0001-9227-5949,sname='Espaillat']{Catherine C. Espaillat} 
\affiliation{Institute for Astrophysical Research, Boston University, 725 Commonwealth Avenue, Boston, MA 02215, USA}
\email{cce@bu.edu}


\begin{abstract}
Isotopic abundance ratios in protoplanetary disks are critical for understanding volatile inheritance and chemical evolution in planet-forming environments. We present Atacama Large Millimeter/submillimeter Array observations of the rare isotopologue $^{13}$C$^{18}$O(2--1) at $\sim$0\farcs3 resolution from the disk around the Herbig Ae star HD~163296, combined with archival observations of C$^{17}$O(2--1), C$^{18}$O(1--0), and C$^{17}$O(1--0), to empirically constrain carbon and oxygen isotopic ratios without detailed disk modeling. Both the C$^{17}$O/$^{13}$C$^{18}$O(2--1) and C$^{18}$O/C$^{17}$O(1--0) flux ratios rise sharply across the CO snowline and flatten beyond 1\farcs5 ($r \gtrsim 150$\,au), where the emission becomes optically thin. This transition, reflecting a steep drop in CO column density set by the disk’s thermal structure, makes HD~163296 an optimal case for isotopic analysis. Using beam-averaged intensities of the four transitions measured in this optically thin region, we derive isotopic ratios of $^{12}$C/$^{13}$C = $75.3^{+14.7}_{-11.4}$ and $^{18}$O/$^{17}$O = $3.28^{+0.31}_{-0.26}$, both consistent with local interstellar medium values. The $^{16}$O/$^{18}$O ratio remains weakly constrained due to moderate optical depth in the C$^{18}$O(1--0) line and degeneracy with CO column density. These results demonstrate that rare CO isotopologues can provide robust, empirical constraints on isotopic ratios in disks when sharp structural transitions allow for the identification of optically thin regions, and establish HD~163296 as a benchmark for extending such studies to other systems with resolved snowline structures.

\end{abstract}



\section{Introduction}
Isotopic abundance ratios in disk gases provide key clues about the chemical heritage and evolution of planetary systems. 
In particular, the carbon ($^{12}$C/$^{13}$C) and oxygen ($^{18}$O/$^{17}$O) isotope ratios link protoplanetary disks to their 
parent molecular clouds and to solar system material. 
The local interstellar medium (ISM) exhibits a carbon isotope ratio significantly lower than the modern solar value 
(the ISM $^{12}$C/$^{13}$C $\approx$ 68--70 vs. $\approx$ 89 in terrestrial material; e.g., \citealt{Milam2005}), 
reflecting Galactic chemical evolution. 
Measurements of CN in comets show $^{12}$C/$^{13}$C $\sim$ 91$\pm$4 \citep{Manfroid2009}, suggesting that carbon in cometary ices, and by extension in the protosolar disk, was not strongly fractionated compared to Earth.  
Protoplanetary disks, which mediate the inheritance or resetting of these isotopic ratios, are therefore key to tracing the chemical connection between the ISM and planetary systems.

Recent observations show that isotopic ratios in disks can deviate from ISM values due to disk-specific physical conditions and isotope-selective chemistry that alter CO isotopologue abundances with radius. In TW~Hya, an analysis of optically thin line wings finds $^{12}$CO/$^{13}$CO $\simeq 21 \pm 5$ at $r \sim$70--110\,au, rising to $\gtrsim$84 beyond $\sim$130\,au, revealing strong spatial variations in the carbon isotope ratio \citep{Yoshida2022}. Models that explicitly incorporate isotope-selective chemistry, including selective photodissociation and low-temperature exchange reactions, are essential to reproduce and interpret such spatial variations. Selective photodissociation arises because rare isotopologues such as C$^{18}$O and C$^{17}$O experience reduced self‑ and mutual shielding compared to $^{12}$CO and $^{13}$CO, making them more easily dissociated in UV‑irradiated surface layers or in disks with significant grain growth \citep{Visser2009}. Low‑temperature exchange reactions can further enhance $^{13}$CO in cold, CO‑depleted regions where ion–molecule chemistry is efficient \citep{Furuya2022, Lee2024}. These pathways naturally produce vertical gradients in isotopologue abundances, since photodissociation dominates in the warm surface while exchange reactions operate more effectively in the cold midplane. As a result, midplane‑tracing lines may yield different apparent isotopic ratios than lines that preferentially arise from higher disk layers. Recent thermochemical implementations in DALI and NAUTILUS, which include isotope‑selective photodissociation, grain‑surface CO processing, and CO-–CO$_2$ interconversion, demonstrate that these combined effects generate substantial radial and vertical structure in CO isotopologue abundances and can bias gas‑mass determinations if not accounted for \citep{Miotello2016, Miotello2021, Ruaud2022, Deng2023, Deng2025}. For oxygen, optically thin HCO$^+$ isotopologue ratios toward TW~Hya yield $^{13}$CO/C$^{18}$O = $8.3 \pm 2.6$, consistent with the local ISM \citep{Furuya2022}, suggesting no strong evidence for $^{18}$O depletion in the bulk CO gas.

Other carbon-bearing molecules in disks show different isotopic behaviors. \citet{Hily-Blant2019} reported that the disk-averaged H$^{12}$CN/H$^{13}$CN ratio in TW~Hya is $86 \pm 4$, higher than the local ISM value and closer to the solar value. In contrast, \citet{Yoshida2024} measured the $^{12}$CN/$^{13}$CN ratio to be $70^{+9}_{-6}$ at 30--80\,au, consistent with the ISM, but differing from the ratios found for CO and HCN in the same disk. These results highlight that different molecules can trace distinct reservoirs of carbon and may carry different isotopic signatures, underscoring the complexity of disk chemistry and the importance of molecule-specific diagnostics.

Accurate determinations of C and O isotopic ratios in disks are needed for understanding disk chemistry and for comparison to solar system materials. From the observational perspective, there are two key challenges: optical depth and sensitivity. 
First, the abundant CO isotopologue lines are often optically thick, and therefore, the line emission does not directly correspond to 
column density. While rare species like C$^{18}$O are typically assumed to be optically thin, recent studies demonstrate that 
C$^{18}$O (and even C$^{17}$O) can become optically thick in bright disks \citep{Law2021}, 
leading to biased abundance estimates if optical depth is not properly accounted for. 
The optical depth of CO isotopologue lines depends sensitively on the disk’s vertical temperature structure. 
As shown by \citet{Qi2024}, disks with a thick vertically isothermal region around the midplane (VIRaM layer) exhibit a sharp drop in CO column density across the CO snow line, which provides a well-defined boundary where emission transitions from optically thick to 
optically thin. In contrast, disks with a thin VIRaM layer show a more gradual radial decrease in CO column density, making it difficult 
to identify the optically thin region observationally. Second, although the outermost disk regions eventually reach low enough 
column densities to ensure low optical depth emission from the rare, optically thin CO isotopologues, their emission is 
intrinsically weak, often precluding robust detections with current instruments and limiting isotopic ratio measurements. 

Given the challenges of disentangling optical depth effects from underlying disk structure when interpreting CO isotopologue emission, it is essential to develop observational strategies that enable robust and preferably model-independent determinations. Measuring flux ratios of multiple rare CO isotopologues observed at the same excitation level, such as C$^{17}$O and C$^{18}$O in the $J=1\rightarrow0$ transition or $^{13}$C$^{18}$O and C$^{17}$O in the $J=2\rightarrow1$ transition, provides an empirical means to constrain isotopic ratios while minimizing assumptions. This approach parallels hyperfine analyses of molecules like CN and HCN \citep[e.g.,][]{Hily-Blant2019}, but is here applied to CO isotopologues. The present study is the first to use high signal-to-noise $^{13}$C$^{18}$O(2--1) observations of the HD~163296 disk to empirically identify an optically thin region beyond the sharply defined CO snowline, directly from the flux ratio behavior, without relying on detailed physical or chemical modeling. This enables more direct and model-independent determinations of the $^{12}$C/$^{13}$C and $^{18}$O/$^{17}$O ratios in a protoplanetary disk. 
The methodology established here provides a framework for extending isotopic studies to other systems with resolved snowline structures.

Our letter is organized as follows. In Section~\ref{sec:obs} we describe the Atacama Large Millimeter/submillimeter Array (ALMA) observations and present the radial profiles of the CO isotopologues, highlighting the sharp CO snowline transition. Then, we quantify how the same transition flux ratios trace optical depth and apply this diagnostic to HD~163296. 
In Section~\ref{sec:mcmc} we use beam-averaged line intensities in the optically thin outer disk to constrain isotopic ratios through Markov Chain Monte Carlo (MCMC) fitting. 
In Section~\ref{sec:conclusions} we summarize our conclusions and discuss the implications of our results for disk chemistry and comparisons with other disks and solar system reservoirs. 

\section{Observations and Results}\label{sec:obs}

The disk around the Herbig Ae star HD~163296 provides an optimal environment to search for optically thin regions in rare CO isotopologue emission, owing to its previously established sharp CO snowline transition at a readily accessible spatial scale \citep{Qi2024}. New, deep ALMA observations of the $^{13}$C$^{18}$O(2--1) line provide the basis for a fresh examination. A detailed description of the observations (Project ID: 2021.1.00899.S, PI: K. Zhang) is given by T. Armitage et al. (ApJ accepted).
The data reduction and imaging were performed using \texttt{CASA} v6.6.3 \citep{McMullin2007,CASA+2022}, following standard pipeline calibration, self-calibration for improved signal-to-noise, and imaging with the \texttt{tclean} task using Briggs weighting (robust 0.5). 
The final image has a beam size of $0\farcs28 \times 0\farcs25$, CLEANed to a $3\sigma$ threshold, corrected for the so called \citet[; JvM]{Jorsater1995} effect \citep[see also][]{Czekala2021}, and primary beam corrected. The velocity field of the $^{13}$C$^{18}$O(2--1) emission is consistent with Keplerian rotation in the disk and matches the kinematic pattern traced by the other CO isotopologues (see Appendix~\ref{app:13c18o_mom1}).

Figure~\ref{fig:13c18o21_profile} shows the radial profile of the $^{13}$C$^{18}$O(2--1) integrated intensity and its radial derivative. The intensity decreases smoothly with radius, and the derivative exhibits a clear local minimum at 0\farcs75, coincident with the CO snowline radius inferred from other CO isotopologues \citep{Qi2024}. This feature indicates a sharp drop in the CO column across the snowline. At this location, the midplane temperature is approximately 20\,K, near the CO freezeout threshold; the $^{13}$C$^{18}$O(2--1) transition has $E_u/k \approx 15$\,K and thus traces gas close to this thermal boundary.
\citet{Qi2024} performed forward radiative‐transfer calculations for the C$^{17}$O(2–1), C$^{18}$O(1–0), and C$^{17}$O(1–0) lines and showed that the observed dip in the radial derivative of the integrated-intensity profile cannot be produced by plausible excitation gradients alone and  a sharp drop in the CO column density at the snowline is required. Because $^{13}$C$^{18}$O(2--1) is optically thinner than those transitions, the persistence of the derivative dip in this line further indicates that the feature is not an opacity artifact. In combination with the archival C$^{17}$O and C$^{18}$O data, the new $^{13}$C$^{18}$O observations allow us to constrain the region where all lines are optically thin, enabling a direct empirical determination of the isotopic ratios.

\begin{figure}[ht!]
\centering
\includegraphics[width=\linewidth]{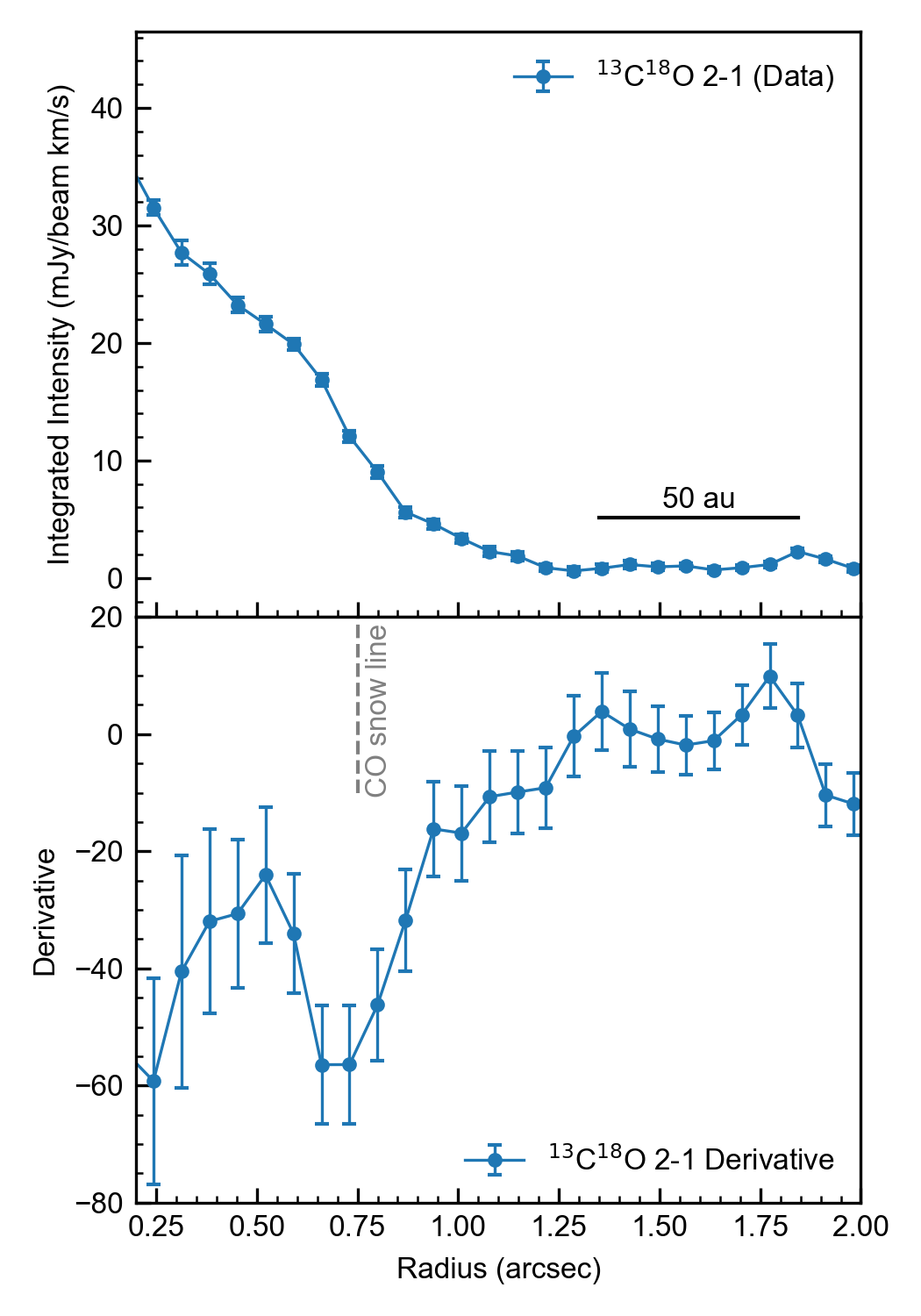}
\caption{Radial profile (top) and derivative (bottom) of the $^{13}$C$^{18}$O(2--1) integrated intensity toward HD~163296. The derivative profile shows a clear local minimum at $\sim$0.75\arcsec, consistent with the location of the sharp CO snowline transition observed in C$^{17}$O and C$^{18}$O lines \citep{Qi2024}.}
\label{fig:13c18o21_profile}
\end{figure}

\begin{figure*}[ht!]
\centering
\includegraphics[width=0.49\linewidth]{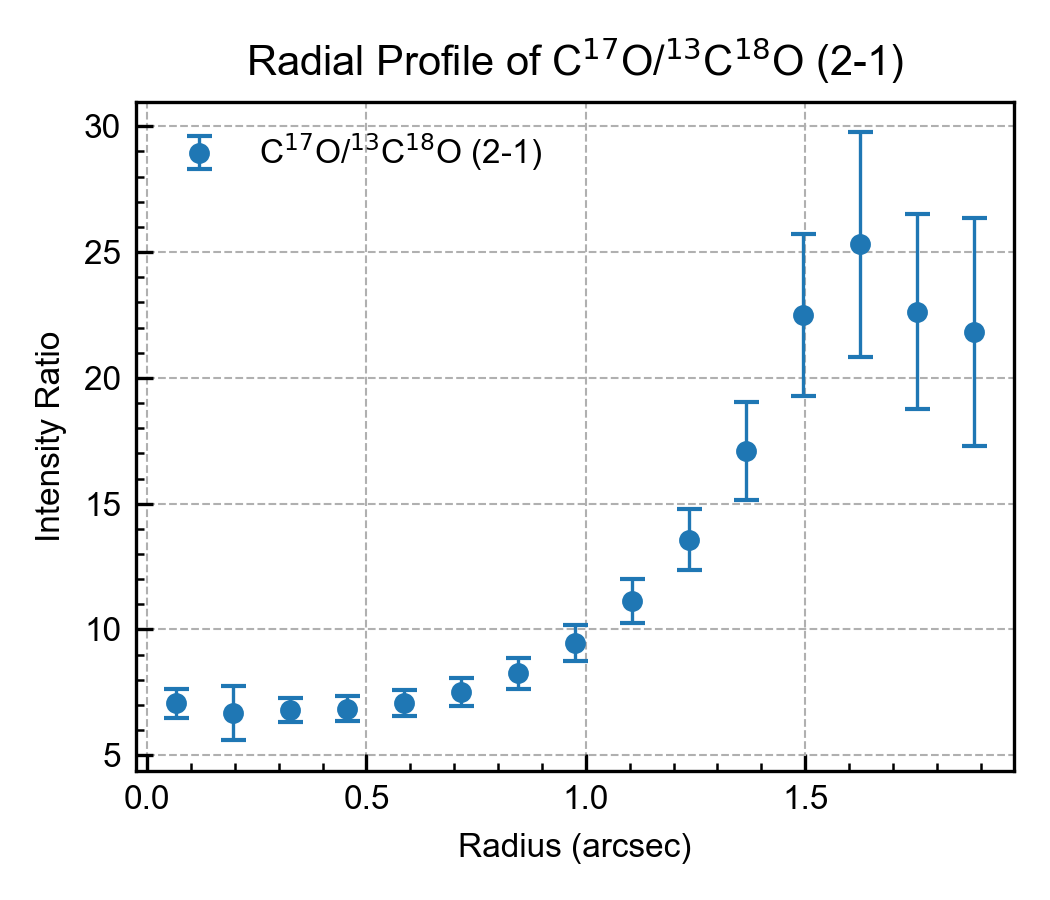}
\includegraphics[width=0.49\linewidth]{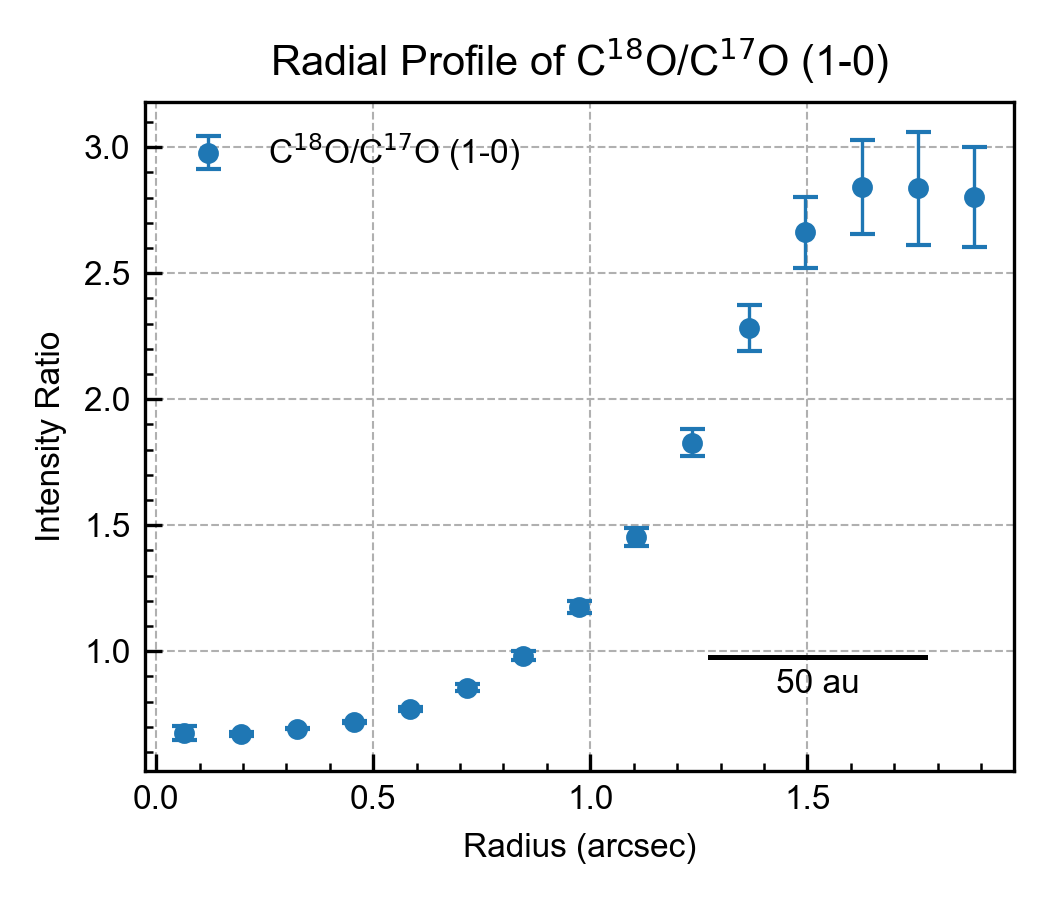}
\caption{Radial profiles of the same-transition isotopologue intensity ratios. 
\textit{Left:} C$^{17}$O/$^{13}$C$^{18}$O(2--1) ratio, flat at small radii, rising steeply across the CO snowline, and converging to $\sim$20 beyond 1\farcs5. Uncertainties include both statistical errors and a 7.1\% systematic calibration term, as the two lines were observed in different bands and epochs. 
\textit{Right:} C$^{18}$O/C$^{17}$O(1--0) ratio, showing similar radial behavior and approaching $\sim$2.7 beyond 1\farcs5. Only statistical uncertainties are shown, since both lines were observed in the same band and share a common amplitude calibration. 
In both panels, the rapid increase followed by a flat plateau indicates the sharp decline in CO column density across the snowline and identifies the outer disk as an optically thin region suitable for isotopic ratio analysis.}
\label{fig:ratio_profiles}
\end{figure*}

To enable accurate flux ratio measurements, we reimaged each transition to a common circularized beam of $0\farcs52$, defined by the major axis of the synthesized beam for C$^{17}$O(2--1) with robust 0.5 weighting and a $u,v$ taper, following the MAPS cleaning strategy \citep[see][]{Czekala2021}. 
The JvM correction and primary beam correction were applied consistently to all transitions, and deprojected, azimuthally averaged radial profiles were extracted using the \texttt{radial\_profile} function from \texttt{GoFish} \citep{Teague2019}.
Figure~\ref{fig:ratio_profiles} presents the resulting radial profiles of the C$^{17}$O/$^{13}$C$^{18}$O(2--1) and C$^{18}$O/C$^{17}$O(1--0) flux ratios.
For the C$^{17}$O/$^{13}$C$^{18}$O(2--1) profile, we propagate both the statistical map uncertainties and a 7.1\% systematic calibration term that represents the worst case of uncorrelated 5\% absolute flux errors between the Band~6 datasets, whereas for the C$^{18}$O/C$^{17}$O(1--0) profile we include only statistical uncertainties because the two Band~3 lines share a common flux scale and their calibration errors are highly correlated. A detailed discussion of flux calibration uncertainty propagation for these ratios is given in Appendix~\ref{app:uncertainty_ratios}.

Both ratios remain relatively flat inside 0\farcs75, consistent with high optical depth, then rise rapidly across the CO snowline and converge to plateaus beyond 1\farcs5 ($r \gtrsim 150$\,au), where all lines become optically thin. This rapid increase, followed by a common plateau, reflects the steep decline in CO column density across the CO snowline \citep{Qi2024}. Because the intrinsic snowline transition is demonstrated to be sharp, the $0\farcs52$ beam broadens the observed rise in the flux ratios over a radial range of approximately 0\farcs5. This modest beam smearing smooths the slope of the transition but does not affect the optically thin plateau values that are used for the isotopic ratio analysis.
The C$^{17}$O/$^{13}$C$^{18}$O(2--1) ratio approaches approximately 20, consistent with the ISM value expected under optically thin emission conditions for both lines, while the C$^{18}$O/C$^{17}$O(1--0) ratio stabilizes near 2.7, implying moderate optical depth in the C$^{18}$O(1--0) line for ISM-like abundances (see Section~\ref{sec:ratios}).  Together, these behaviors identify the outer region around and beyond 1\farcs5 where isotopic ratios can be reliably constrained. 

\subsection{Constraining Optical Depths with Same-transition Isotopologue Flux Ratios}\label{sec:ratios}
To quantify how same-transition isotopologue flux ratios vary with line-center optical depth, we performed a systematic non-LTE radiative transfer analysis using the \code{RADEX} code \citep{vanderTak2007} through the Python package \code{SpectralRadex} \citep{Holdship2021}.
We adopted molecular data from the LAMDA database \citep{Schoier2005}, using collisional rate coefficients from \citet{Yang2010}. Collisions with both ortho- and para-H$_2$ were included, assuming an ortho-to-para ratio of 3:1, and level populations were solved with the Cosmic Microwave Background (CMB) as the only background radiation field ($T_{\rm bg} = 2.73$\,K).  We adopted canonical ISM isotopic ratios \citep{Wilson1999}: $^{12}$C/$^{13}$C$ = 69$, $^{16}$O/$^{18}$O$ = 557$, and $^{18}$O/$^{17}$O$ = 3.6$, and explored a grid of kinetic temperatures ($T_{\rm kin} = 30$, 50, 70\,K), molecular hydrogen densities ($n_{\rm H_2} = 10^5$--$10^8$\,cm$^{-3}$), and CO column densities ($\log N_{\rm CO} = 16$--19).

\begin{figure*}[ht!]
\centering
\includegraphics[width=\linewidth]{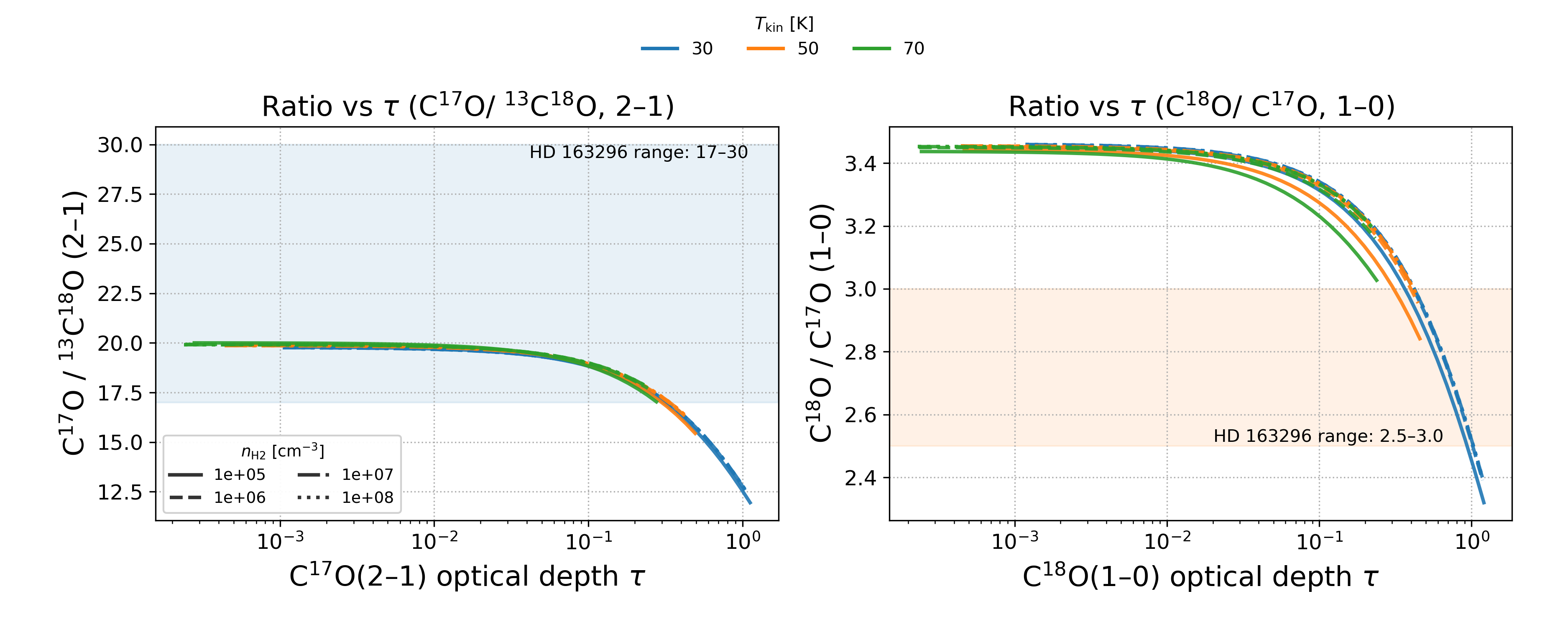}
\caption{\code{RADEX}-calculated same-transition isotopologue flux ratios as a function of line-center optical depth for a range of $n_{\rm H_2}$ and $T_{\rm kin}$. 
Upper: C$^{17}$O/$^{13}$C$^{18}$O(2--1) ratio versus C$^{17}$O(2--1) optical depth. 
Lower: C$^{18}$O/C$^{17}$O(1--0) ratio versus C$^{18}$O(1--0) optical depth. 
In the optically thin regime ($\tau \ll 1$), the ratios converge to the values set by ISM isotopic abundances (19.2 and 3.6, respectively), and decline rapidly once the more abundant line becomes optically thick ($\tau \gtrsim 1$). 
The shaded regions indicate the plateau ranges of the ratios measured in HD~163296 beyond 1\farcs5 (see Figure~\ref{fig:ratio_profiles}), where all transitions are optically thin. 
This comparison shows that the observed plateau values fall squarely in the optically thin regime, confirming their suitability for constraining isotopic ratios.}
\label{fig:ratio_vs_tau}
\end{figure*}

At each grid point, we derived the isotopologue column densities for C$^{17}$O, C$^{18}$O, and $^{13}$C$^{18}$O from the adopted abundance ratios, then used \code{RADEX} to compute the line-center optical depth $\tau$ and integrated line flux (in K\,km\,s$^{-1}$) for the primary transition (either C$^{18}$O(1--0) or C$^{17}$O(2--1)). The flux ratio was then calculated relative to its rarer counterpart, $^{13}$C$^{18}$O(2--1) or C$^{17}$O(1--0), respectively. Figure~\ref{fig:ratio_vs_tau} shows the flux ratios as a function of line-center optical depth. 
In the optically thin regime ($\tau \ll 1$), the flux ratios converge to the values set by the adopted ISM isotopic abundances: 19.2 for C$^{17}$O/$^{13}$C$^{18}$O(2--1), corresponding to $^{12}$C/$^{13}$C divided by $^{18}$O/$^{17}$O ($69/3.6$) and 3.6 for C$^{18}$O/C$^{17}$O(1--0), reflecting the $^{18}$O/$^{17}$O ratio.
We explicitly tested the sensitivity of these ratio-$\tau$ curves to $\pm30$\% variations in the assumed isotopic ratios (Appendix~\ref{app:sensitivity}) and found that while the thin-limit plateau shifts as expected, the shape of the ratio as a function of $\tau$ over the $\tau \lesssim 0.3$ regime used to identify the optically thin zone in HD~163296 is essentially unchanged.
As the optical depth increases ($\tau \gtrsim 1$), the more abundant line saturates, and the ratio declines rapidly, 
largely independent of temperature or density across the disk-relevant range. Because both lines in each pair arise from the same $J \rightarrow J-1$ transition and have nearly identical frequencies, their ratios substantially reduce sensitivity to excitation temperature and beam dilution, though vertical differences in the emitting layers can introduce residual effects.

These theoretical expectations match our observations of HD~163296: the measured ratios in Figure~\ref{fig:ratio_profiles} rise steeply across the snowline and then level off beyond 1\farcs5. This agreement demonstrates that the ratios provide a direct diagnostic of optical depth and confirms that the outer disk of HD~163296 offers a robust optically thin region suitable for empirical isotopic ratio determinations.

\section{MCMC Fitting to Constrain Isotopic Ratios} \label{sec:mcmc}
As described in Section~\ref{sec:obs}, all CO isotopologue data were reimaged to a common circularized beam of $0\farcs52$ to enable accurate flux ratio measurements. This common beam is essential for consistent comparison across transitions. We then used the \texttt{radial\_profile} function from the \texttt{GoFish} package \citep{Teague2019} to extract deprojected, azimuthally averaged radial intensity profiles. To constrain the isotopic ratios in the optically thin region, we retrieved the integrated intensity in an annulus centered at 1\farcs5, which is effectively smoothed over the $0\farcs52$ beam. These beam-averaged values are summarized in Table~\ref{tab:fluxes} and per–channel rms values at the working spectral resolutions are listed in Appendix~\ref{app:line-noise}, Table~\ref{tab:line_noise}. 

\begin{table}[ht!]
\centering
\caption{Beam-averaged CO isotopologue integrated intensities at 1\farcs5 (around 150\,au) in HD~163296}
\label{tab:fluxes}
\begin{tabular}{lcc}
\hline
Transition & Flux (mJy\,beam$^{-1}$\,km\,s$^{-1}$) & Flux (K\,km\,s$^{-1}$) \\
\hline
C$^{17}$O(1--0) & $5.82 \pm 0.29$ & $2.09 \pm 0.10$ \\
C$^{18}$O(1--0) & $15.49 \pm 0.25$ & $5.80 \pm 0.09$ \\
$^{13}$C$^{18}$O(2--1) & $3.33 \pm 0.41$ & $0.34 \pm 0.04$ \\
C$^{17}$O(2--1) & $74.85 \pm 1.17$ & $6.70 \pm 0.10$ \\
\hline
\end{tabular}
\tablecomments{Quoted uncertainties are statistical (map) errors. A nominal 5\% ALMA absolute flux calibration uncertainty applies to each line and should be added in quadrature when using absolute intensities. Values correspond to integrated intensities within the $0\farcs52$ common beam centered at 1\farcs5.}
\end{table}

Because the optically thin area is confined to the outer disk and the rare isotopologue lines are weak, we adopted this single beam-averaged annulus at 1\farcs5 for quantitative fitting. Within the uncertainties, no significant radial trend is detected in either ratio across the outer disk region (Fig.~\ref{fig:ratio_profiles}), so a single representative radius is sufficient for the isotopic analysis.

We used an MCMC approach to fit these intensities using \code{RADEX}
with seven free parameters: kinetic temperature ($T_{\rm kin}$), molecular hydrogen density ($n_{\rm H_2}$), 
CO column density ($N_{\rm CO}$), line width ($\Delta V$), and three isotopic ratios ($^{12}$C/$^{13}$C, $^{16}$O/$^{18}$O, and $^{18}$O/$^{17}$O). 
Priors were chosen to be uninformative but physically bounded: we adopted uniform priors in linear space for
$T_{\rm kin}\!\sim\!{\cal U}(10,200)$\,K and $\Delta V\!\sim\!{\cal U}(0.1,0.6)$\,km\,s$^{-1}$, and uniform-in-log priors for the scale parameters
$\log_{10} n_{\rm H_2}\!\sim\!{\cal U}(4,9)$ and $\log_{10} N_{\rm CO}\!\sim\!{\cal U}(16,20)$.
For the isotopic ratios, we used broad uniform priors,
$^{12}$C/$^{13}$C$\!\sim\!{\cal U}(10,200)$,
$^{16}$O/$^{18}$O$\!\sim\!{\cal U}(100,1000)$,
and $^{18}$O/$^{17}$O$\!\sim\!{\cal U}(1,10)$,
so that the posteriors are driven by the data rather than by the priors.
Table~\ref{tab:mcmc_results} summarizes the median posterior values and their 68\% confidence intervals.

\begin{table}[ht!]
\centering
\caption{MCMC Parameter Estimates at 1\farcs5 (around 150\,au)}
\label{tab:mcmc_results}
\begin{tabular}{lc}
\hline
Parameter & Value \\
\hline
$T_{\rm kin}$ (K) & $112.1^{+50.5}_{-38.3}$ \\
$\log_{10}(n_{\rm H_2}$/cm$^{-3}$) & $6.05^{+1.34}_{-1.26}$ \\
$\log_{10}(N_{\rm CO}$/cm$^{-2}$) & $19.25^{+0.25}_{-0.32}$ \\
$\Delta V$ (km\,s$^{-1}$) & $0.32^{+0.18}_{-0.15}$ \\
$^{12}$C/$^{13}$C & $75.3^{+14.7}_{-11.4}$ \\
$^{16}$O/$^{18}$O & $599^{+276}_{-310}$ \\
$^{18}$O/$^{17}$O & $3.28^{+0.31}_{-0.26} $ \\
\hline
\end{tabular}
\end{table}

The best-fit isotopic ratios at 1\farcs5 and comparisons with ISM values \citep{Wilson1999} are as follows:
\begin{itemize}
    \item $^{12}$C/$^{13}$C = $75.3^{+14.7}_{-11.4}$, consistent with the ISM value of $69 \pm 6$;
    \item $^{18}$O/$^{17}$O = $3.28^{+0.31}_{-0.26}$, consistent with the ISM value of $3.6 \pm 0.2$;
    \item $^{16}$O/$^{18}$O = $599^{+276}_{-310}$, consistent with the ISM value of $557 \pm 30$ within large uncertainties.
\end{itemize}

These results confirm that, in the optically thin outer disk of HD~163296, the carbon and oxygen isotopic ratios are broadly consistent with those of the local ISM. The corner plot of the posterior distributions, which illustrates parameter covariances and uncertainties, is shown in Appendix~\ref{app:posterior}. As demonstrated by the calibration sensitivity tests in Appendix~\ref{app:calib}, $5$--$10$\% flux scale offsets do not shift the inferred isotopic ratios beyond their quoted uncertainties. The $^{12}$C/$^{13}$C and $^{18}$O/$^{17}$O ratios show minimal covariance with the physical parameters, indicating that they are constrained directly by the $^{13}$C$^{18}$O and C$^{17}$O fluxes in the optically thin regime. The relatively clean constraints on these ratios underscore the value of rare isotopologues for probing intrinsic elemental abundances in disks. In contrast, the $^{16}$O/$^{18}$O ratio remains weakly constrained because C$^{18}$O(1--0) is only marginally optically thick at line center and because of strong parameter covariance. In particular, $^{16}$O/$^{18}$O correlates positively with $\log N_{\rm CO}$: higher oxygen isotope ratios can be offset by lower CO columns while preserving the C$^{18}$O intensity. The $T_{\rm kin}$, $n_{\rm H_2}$, and $\Delta V$ parameters also show broad, unconstrained posteriors, as expected for optically thin conditions in which line intensities are relatively insensitive to these quantities. We note that in our \code{RADEX} calculations, $\Delta V$ represents the local thermal plus microturbulent FWHM, not the beam-smeared Keplerian shear. The posterior value $\Delta V\sim0.3$~km s$^{-1}$ therefore reflects local broadening at the emitting layer rather than the total observed profile width.

\section{Conclusions and future work}\label{sec:conclusions}
We have shown that same-transition isotopologue flux ratios provide a direct and model-independent diagnostic of optical depth in protoplanetary disks. New ALMA observations of the rare $^{13}$C$^{18}$O(2--1) line confirm the sharp CO snowline transition in the HD~163296 disk at 0\farcs75. Together with archival C$^{17}$O and C$^{18}$O data reimaged to a common circularized beam, these observations reveal flux ratios that rise steeply across the snowline and converge to constant values beyond 1\farcs5, identifying an outer region where all lines are optically thin. 
The beam-averaged line intensities from this region were used in an MCMC analysis to derive isotopic ratios. The results yield $^{12}$C/$^{13}$C and $^{18}$O/$^{17}$O values consistent with local ISM abundances, while $^{16}$O/$^{18}$O remains poorly constrained due to residual optical depth in the C$^{18}$O(1--0) line. These findings demonstrate that rare CO isotopologues provide robust, empirical constraints on isotopic ratios in disks and establish a framework for extending such studies to other systems with resolved snowline structures.

The ISM-like $^{12}$C/$^{13}$C and $^{18}$O/$^{17}$O ratios in HD~163296’s outer disk suggest that the bulk gas reservoir has remained largely unprocessed since formation. Comparable results have been reported in other disks, such as PDS~70 \citep{Rampinelli2025}, whereas TW~Hya shows strong CO fractionation in the inner disk but near-ISM ratios at larger radii \citep[e.g.,][]{Hily-Blant2019,Yoshida2022}. These comparisons indicate that isotope-selective processes, while present in localized regions or specific molecules, may not strongly alter the outer disk gas, or that efficient mixing with unfractionated material preserves ISM-like compositions. The preservation of interstellar carbon and oxygen isotope ratios in HD~163296 parallels the record in primitive solar system reservoirs, where bulk carbon and oxygen remain close to cosmic values, while nitrogen and hydrogen exhibit large anomalies in comets and meteorites \citep[e.g.,][]{Altwegg2019, Rubin2019}. Together, these results support the view that outer disks can serve as relatively pristine isotopic reservoirs, linking the chemistry of the parent cloud to the material incorporated into planetesimals and comets.

The connection to solar system materials is suggestive but necessarily tentative, because our measurements trace gas in the outer disk, whereas meteoritic and cometary values primarily record ices and refractory solids. Gas–-ice exchange, selective photodissociation, low-temperature isotope exchange, and transport across snowlines can decouple gas-phase and solid-phase isotopic signatures. In this context, the ISM-like gas ratios in HD~163296 indicate that the outer disk can preserve interstellar compositions, but do not by themselves establish the isotopic makeup of the ices that seed planetesimals. A broader, method-consistent survey that combines gas and solid tracers will be needed to discriminate inheritance from disk-level resetting.

Deeper integrations and, still rarer, optically thinner tracers will sharpen these constraints and enable radial tests. In particular, spatially resolved measurements of $^{13}$C$^{18}$O(1--0) and $^{13}$C$^{17}$O(1--0), obtained with matched beams and uniform calibration, would reduce covariance with $N_{\rm CO}$ and directly probe the $^{16}$O/$^{18}$O ratio in the thin limit. Higher sensitivity will also permit annulus-by-annulus fits beyond $1\farcs5$ to search for subtle radial trends and vertical gradients via same–transition ratios. Looking ahead, the ngVLA’s superior 3\,mm sensitivity and subarcsecond resolution at the CO isotopologue (1--0) frequencies will map these faint lines over large areas, providing a powerful, model-independent census of carbon and oxygen isotopic ratios across disks.

\begin{acknowledgments}
We thank the anonymous referee for a constructive report, which improved the paper. C.Q. thanks Floris van der Tak and Alice Booth for helpful discussions.
This paper makes use of the following ALMA data: ADS/JAO.ALMA\#2016.1.00884.S, ADS/JAO.ALMA\#2018.1.01055.L, and \\
ADS/JAO.ALMA\#2021.1.00899.S. ALMA is a partnership of ESO (representing its member states), NSF (USA), and NINS (Japan), together with NRC (Canada), NSTC and ASIAA (Taiwan), and KASI (Republic of Korea), in cooperation with the Republic of Chile. The Joint ALMA Observatory is operated by ESO, AUI/NRAO, and NAOJ.
The National Radio Astronomy Observatory is a facility of the National 
Science Foundation operated under a cooperative agreement by Associated Universities, Inc.
\end{acknowledgments}
%
\facilities{ALMA}

\software{Astropy \citep{Astropy2013,Astropy2018,Astropy2022}, \code{bettermoments} \citep{Teague2018}, \code{CASA} \citep{CASA+2022},  \code{emcee} \citep{emcee}, \code{GoFish} \citep{Teague2019}, Matplotlib \citep{Hunter2007}, NumPy \citep{vanderWalt2011}, \code{RADEX} \citep{vanderTak2007}, \code{SpectralRadex} \citep{Holdship2021}.}         


\appendix
\section{Kinematics of $^{13}$C$^{18}$O(2--1) Emission}
\label{app:13c18o_mom1}

\begin{figure}[ht!]
    \centering
    \includegraphics[width=0.5\textwidth]{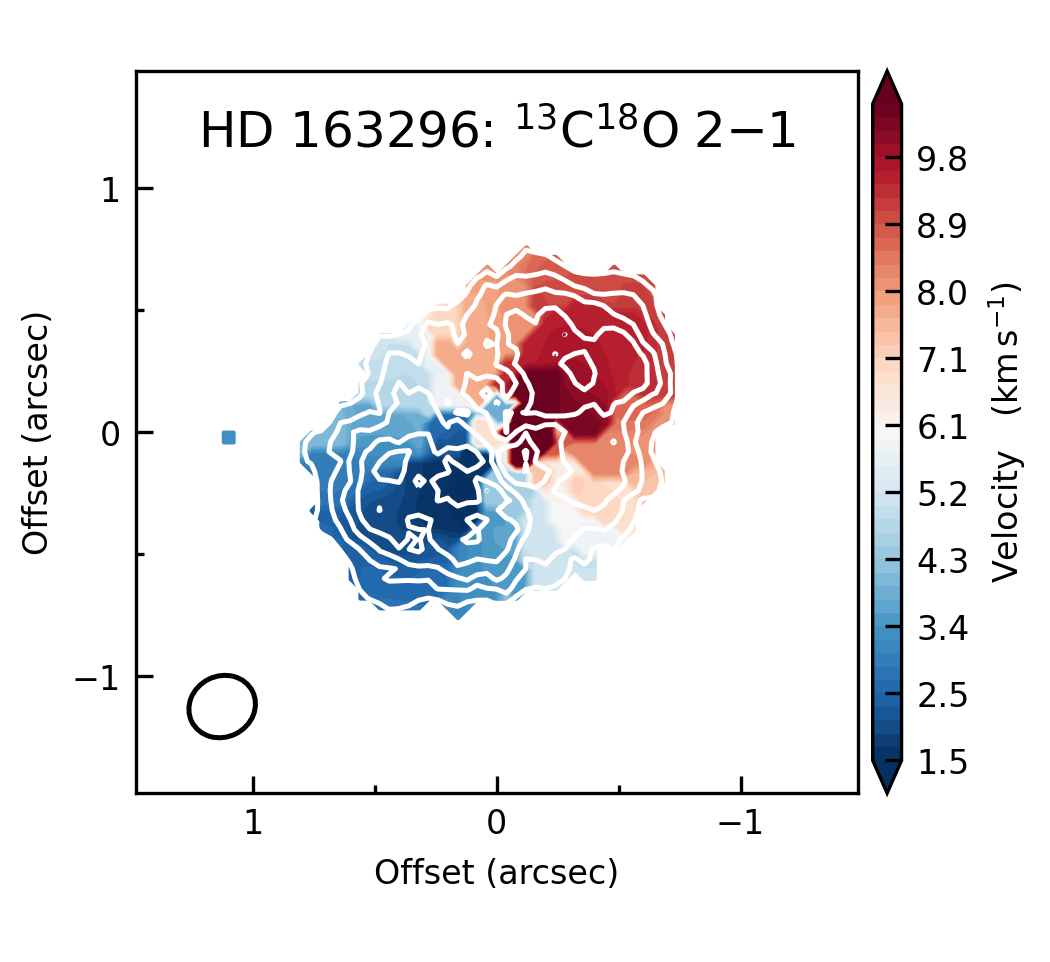}
    \caption{Velocity field (moment-1, color) of the $^{13}$C$^{18}$O $J$=2--1 emission overlaid with white contours of the integrated intensity (moment-0), both derived using the \texttt{bettermoments} package \citep{Teague2018}. The intensity-weighted velocity was calculated with a $2\sigma$ clip to minimize noise bias.
    }
    \label{fig:13c18o_moments}
\end{figure}

To verify that the detected $^{13}$C$^{18}$O(2--1) emission arises from the disk and follows Keplerian rotation, we present the moment-1 (intensity-weighted velocity) map in Figure~\ref{fig:13c18o_moments}, overlaid with moment-0 (integrated intensity) contours. These maps were generated using the \texttt{bettermoments} package \citep{Teague2018}, with a $2\sigma$ intensity clip to mitigate noise bias. The velocity pattern clearly traces a Keplerian gradient aligned with the disk's major axis, and the emission is confined within the disk extent seen in other isotopologues \citep[e.g.,][]{Qi2011}.

\section{Flux Calibration Uncertainty Propagation for Ratios}
\label{app:uncertainty_ratios}

Absolute flux calibration uncertainties introduce systematic errors into line ratio measurements, particularly when the lines are observed in different frequency bands or at different epochs. We adopt the standard ALMA absolute calibration uncertainty of approximately 5\% per dataset. The propagated fractional uncertainty on a flux ratio $R = F_1/F_2$ can be expressed as
\begin{equation}
\frac{\sigma_R}{R} \approx \sqrt{\epsilon_1^2 + \epsilon_2^2 - 2\rho \epsilon_1\epsilon_2},
\end{equation}
where $\epsilon_i$ are the fractional flux calibration uncertainties for the two datasets, and $\rho$ is the correlation coefficient between their amplitude calibration.

For ratios between lines observed in the \textit{same} band and session, such as C$^{18}$O(1--0) and C$^{17}$O(1--0), the calibration errors are highly correlated ($\rho \approx 1$). In this case, the propagated uncertainty from flux calibration is negligible, and the ratio uncertainty is dominated by statistical map noise and small relative effects between spectral windows (typically $\lesssim 1$--2\%).

For ratios involving lines observed in \textit{different} bands or epochs, such as C$^{17}$O(2--1)/$^{13}$C$^{18}$O(2--1), the correlation is lower. In the worst-case scenario where the calibration errors are uncorrelated ($\rho = 0$), the systematic contribution to the ratio uncertainty is $\sqrt{2} \times$ 5\% $\approx$ 7.1\%. We therefore include a 7.1\% systematic calibration uncertainty in quadrature for the C$^{17}$O/$^{13}$C$^{18}$O(2--1) ratio in Figure~\ref{fig:ratio_profiles}.

Beam matching, primary beam correction, and JvM correction are common-mode processing steps applied consistently across all datasets. These steps do not introduce additional systematic uncertainty into the ratio measurements, although minor imperfections in beam circularization or JvM application may contribute small residual effects.

\section{Sensitivity of the Ratio--Optical Depth Diagnostic to Assumed Isotopic Ratios}
\label{app:sensitivity}

We tested whether our optical depth inference depends on the adopted elemental isotope ratios in the \code{RADEX} grids. We focus on the same-transition ratios used in the main text: C$^{17}$O/$^{13}$C$^{18}$O(2--1) and C$^{18}$O/C$^{17}$O(1--0).

\begin{figure*}[t]
\centering
\includegraphics[width=0.9\linewidth]{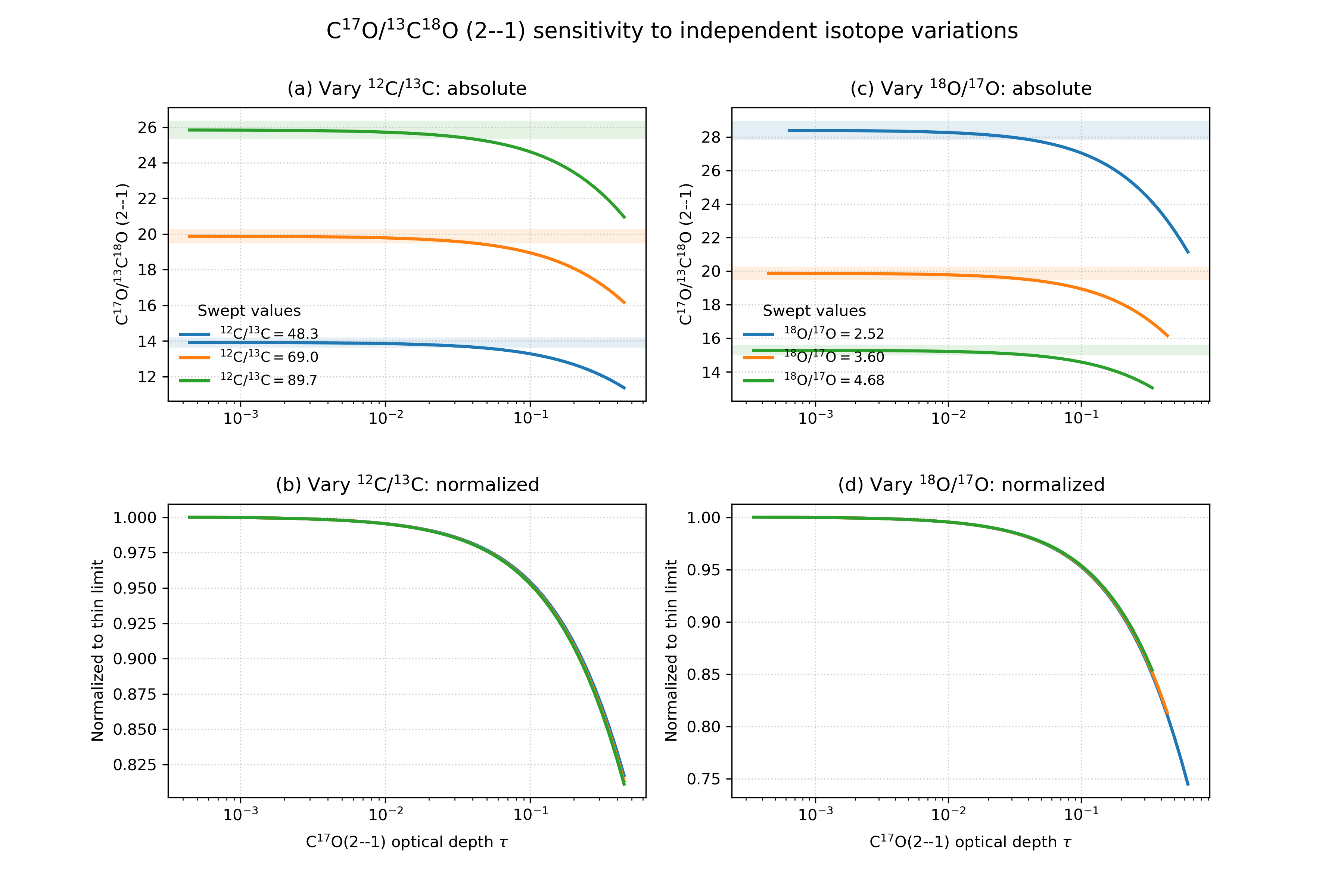}
\caption{
Test A: sensitivity of the C$^{17}$O/$^{13}$C$^{18}$O(2--1) ratio to independent variations in isotopic abundances.
Top left: Absolute ratio as a function of C$^{17}$O(2--1) optical depth when varying $^{12}$C/$^{13}$C by $\pm$30\% at fixed $^{18}$O/$^{17}$O.
Top right: Same, but varying $^{18}$O/$^{17}$O at fixed $^{12}$C/$^{13}$C.
Bottom: All curves normalized by their optically thin limits. The nearly identical shapes confirm that the ratio–$\tau$ response is insensitive to these variations.}
\label{fig:appendix_c17o13c18o_sens}
\end{figure*}

\paragraph{Test A: independent variation of $^{12}$C/$^{13}$C and $^{18}$O/$^{17}$O.}
Figure~\ref{fig:appendix_c17o13c18o_sens} shows how the C$^{17}$O/$^{13}$C$^{18}$O(2--1) ratio changes when $^{12}$C/$^{13}$C is varied by $\pm$30\% at fixed $^{18}$O/$^{17}$O (left), and when $^{18}$O/$^{17}$O is varied by $\pm$30\% at fixed $^{12}$C/$^{13}$C (right). As expected, the optically thin limit shifts with the assumed elemental ratios:
\[
\left.\mathrm{C}^{17}\mathrm{O}/{}^{13}\mathrm{C}^{18}\mathrm{O}\right|_{\tau\rightarrow 0}
= (^{12}\mathrm{C}/^{13}\mathrm{C})/(^{18}\mathrm{O}/^{17}\mathrm{O}).
\]
However, after normalizing each curve by its own thin-limit value, the ratio–$\tau$ profiles collapse to nearly identical shapes. This demonstrates that the optical depth response—the turnover away from the thin-limit plateau—is essentially insensitive to $\pm$30\% variations in either elemental ratio.

\begin{figure*}[t]
\centering
\includegraphics[width=0.9\linewidth]{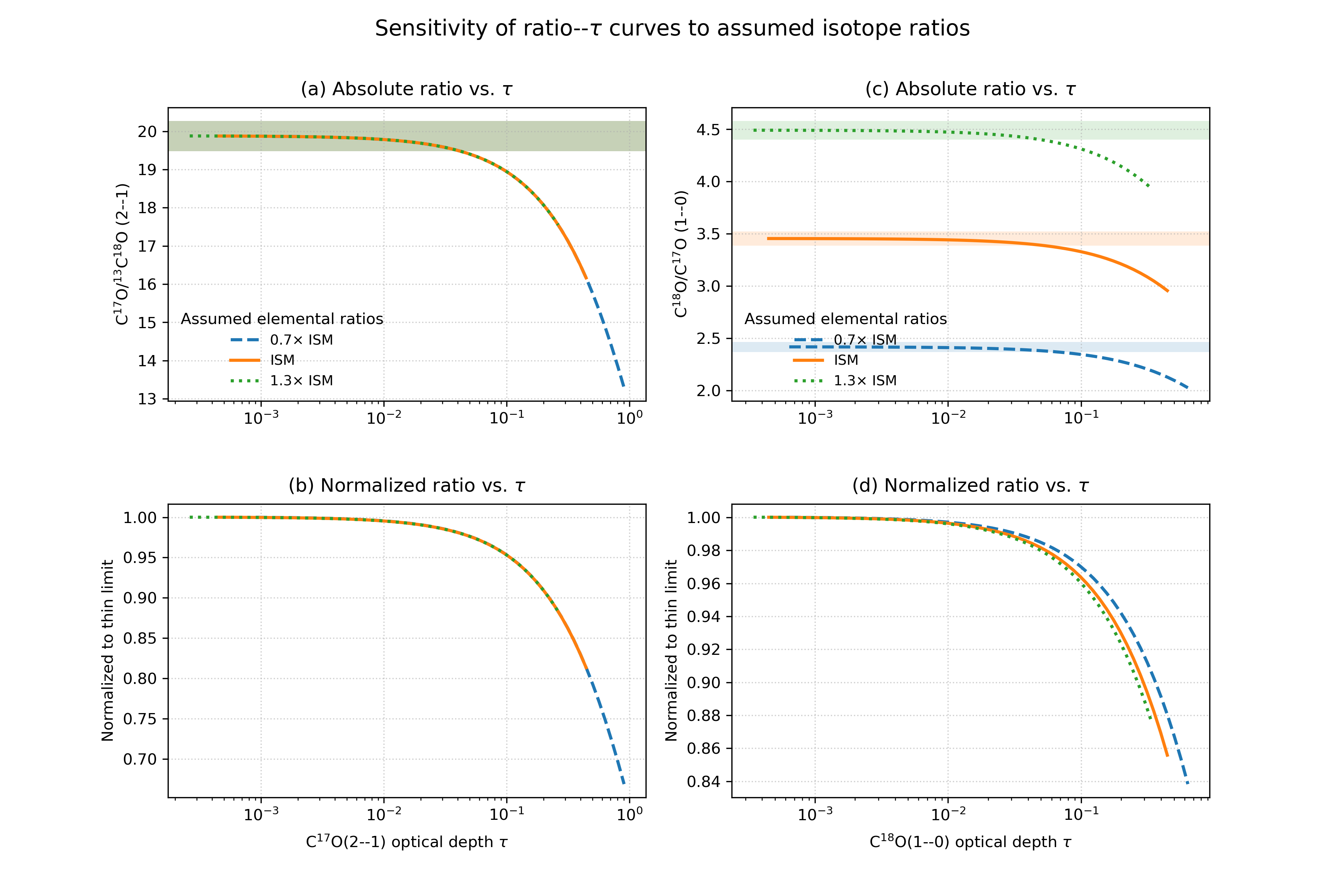}
\caption{
Test B: sensitivity of the ratio–$\tau$ relation to coherent scaling of all isotopic ratios by $\pm$30\% around the ISM values.
Top panels: Absolute C$^{17}$O/$^{13}$C$^{18}$O(2--1) (left) and C$^{18}$O/C$^{17}$O(1--0) (right) ratios as a function of line-center optical depth. 
Bottom panels: Normalized curves collapse to within $\lesssim$1--2\% for $\tau \lesssim 0.3$, confirming the robustness of our optical depth diagnostic and thin-region identification.}
\label{fig:appendix_ratio_tau_sens}
\end{figure*}

\paragraph{Test B: coherent scaling of the full elemental set.}
Figure~\ref{fig:appendix_ratio_tau_sens} shows the effect of scaling all isotopic ratios coherently by $\pm$30\% around the ISM values. Both the C$^{17}$O/$^{13}$C$^{18}$O(2--1) and C$^{18}$O/C$^{17}$O(1--0) ratio profiles are examined. As in Test A, the absolute thin-limit plateaus shift as expected, but the normalized ratio–$\tau$ curves agree to within $\lesssim$1--2\% for $\tau \lesssim 0.3$, the regime used to identify the optically thin region in HD~163296.

\paragraph{Summary.}
Both tests show that while the absolute optically thin limit encodes the elemental isotope ratios, the shape of the ratio–optical depth relation—which we use to locate the optically thin region beyond $1\farcs5$ is robust to plausible ($\pm$30\%) variations in $^{12}$C/$^{13}$C and $^{18}$O/$^{17}$O. Our optical depth mapping and selection of the thin region are therefore not sensitive to the exact isotope ratios assumed in the modeling grid.

\section{Line Properties and Noise}\label{app:line-noise}

Table~\ref{tab:line_noise} summarizes the rest frequencies, spectral channel widths, and per–channel rms sensitivities for the reimaged data products used in this work. All datasets were tapered and Briggs weighted to a common circularized beam (FWHM $0\farcs52$). The quoted $\sigma_{\rm chan}$ values are measured in line–free channels at the tabulated spectral resolutions. 
\begin{table}[ht!]
\centering
\caption{CO isotopologue line properties and per–channel noise (common circularized beam)}
\label{tab:line_noise}
\begin{tabular}{lccc}
\hline
Transition & $\nu_{\rm rest}$ (GHz) & Channel width (km\,s$^{-1}$) & $\sigma_{\rm chan}$ (mJy\,beam$^{-1}$) \\
\hline
C$^{17}$O(1--0)        & 112.359 & 0.50 & 0.51 \\
C$^{18}$O(1--0)        & 109.782 & 0.50 & 0.48 \\
$^{13}$C$^{18}$O(2--1) & 209.419 & 0.32 & 1.17 \\
C$^{17}$O(2--1)        & 224.714 & 0.32 & 2.54 \\
\hline
\end{tabular}
\tablecomments{All datasets share a circularized beam of FWHM $0\farcs52$ (beam position angle PA not applicable). Per–channel rms values are measured in line–free channels after reimaging at the listed spectral resolutions.}
\end{table}

\section{Posterior Structure and Parameter Covariance}
\label{app:posterior}

\begin{figure*}[t]
    \centering
    \includegraphics[width=0.95\textwidth]{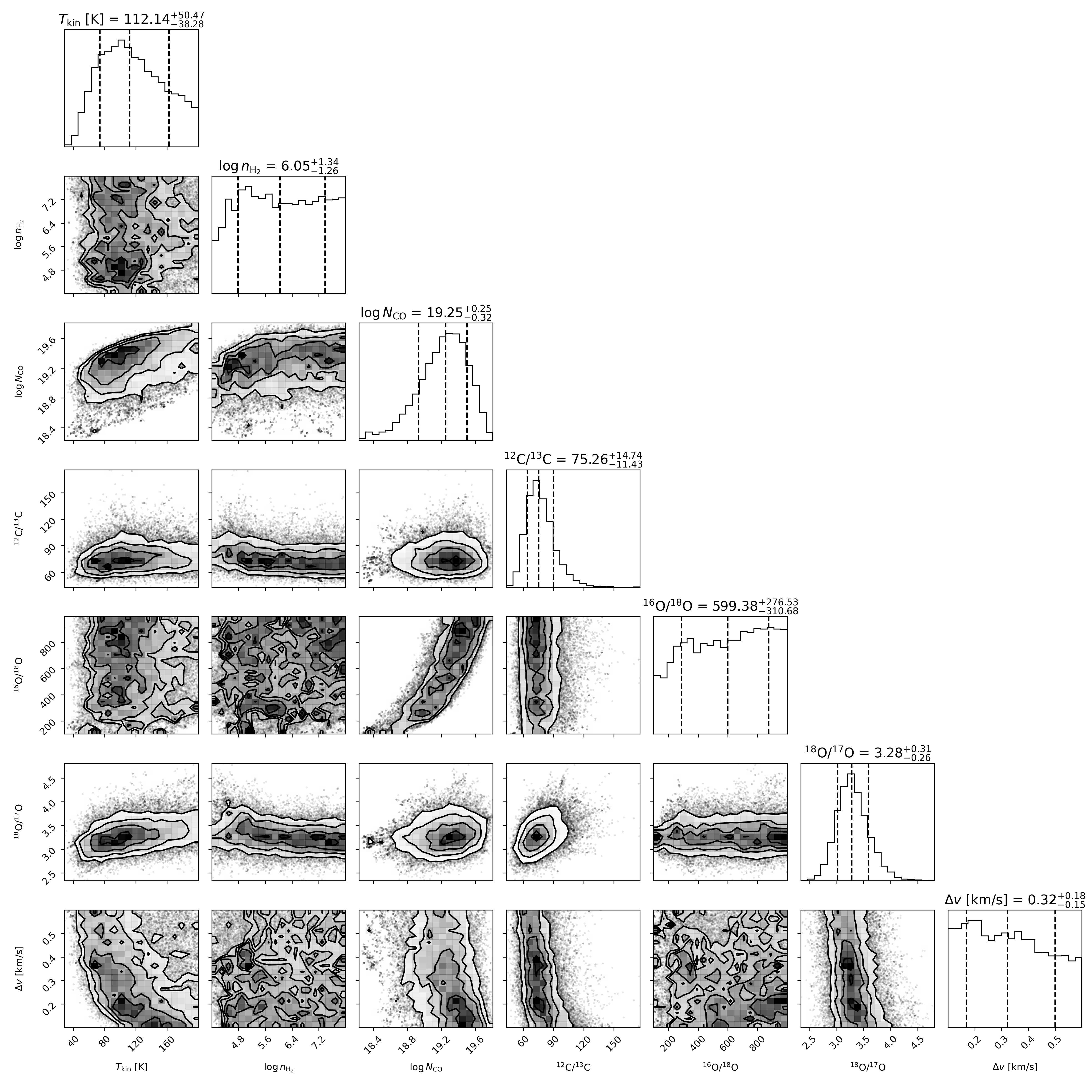}
    \caption{Posterior distributions of the model parameters from the MCMC fit. Diagonal panels show marginalized one–dimensional posteriors; off–diagonal panels show two–dimensional projections that reveal covariances. The $^{12}$C/$^{13}$C and $^{18}$O/$^{17}$O ratios are tightly constrained and largely uncorrelated with other parameters, whereas $^{16}$O/$^{18}$O and $\log N_{\rm CO}$ exhibit strong covariance, reflecting the moderate optical depth of C$^{18}$O(1--0).
    }
    \label{fig:mcmc_corner}
\end{figure*}

\paragraph{Full posterior corner plot.}
Figure~\ref{fig:mcmc_corner} shows the corner plot of the posterior distributions from our MCMC analysis, highlighting covariances between parameters. The marginalized one–dimensional posteriors for $^{12}$C/$^{13}$C and $^{18}$O/$^{17}$O are well converged and approximately Gaussian, yielding precise estimates. In contrast, the posteriors for $^{16}$O/$^{18}$O, $\log N_{\rm CO}$, and to a lesser extent $\Delta V$, are broader and show notable covariance, consistent with the moderate optical depth of C$^{18}$O(1--0) discussed in Section~\ref{sec:mcmc}. These trends also motivate the use of rarer and more optically thin tracers to improve constraints on $^{16}$O/$^{18}$O.

\begin{figure}[t]
    \centering
    \includegraphics[width=0.55\textwidth]{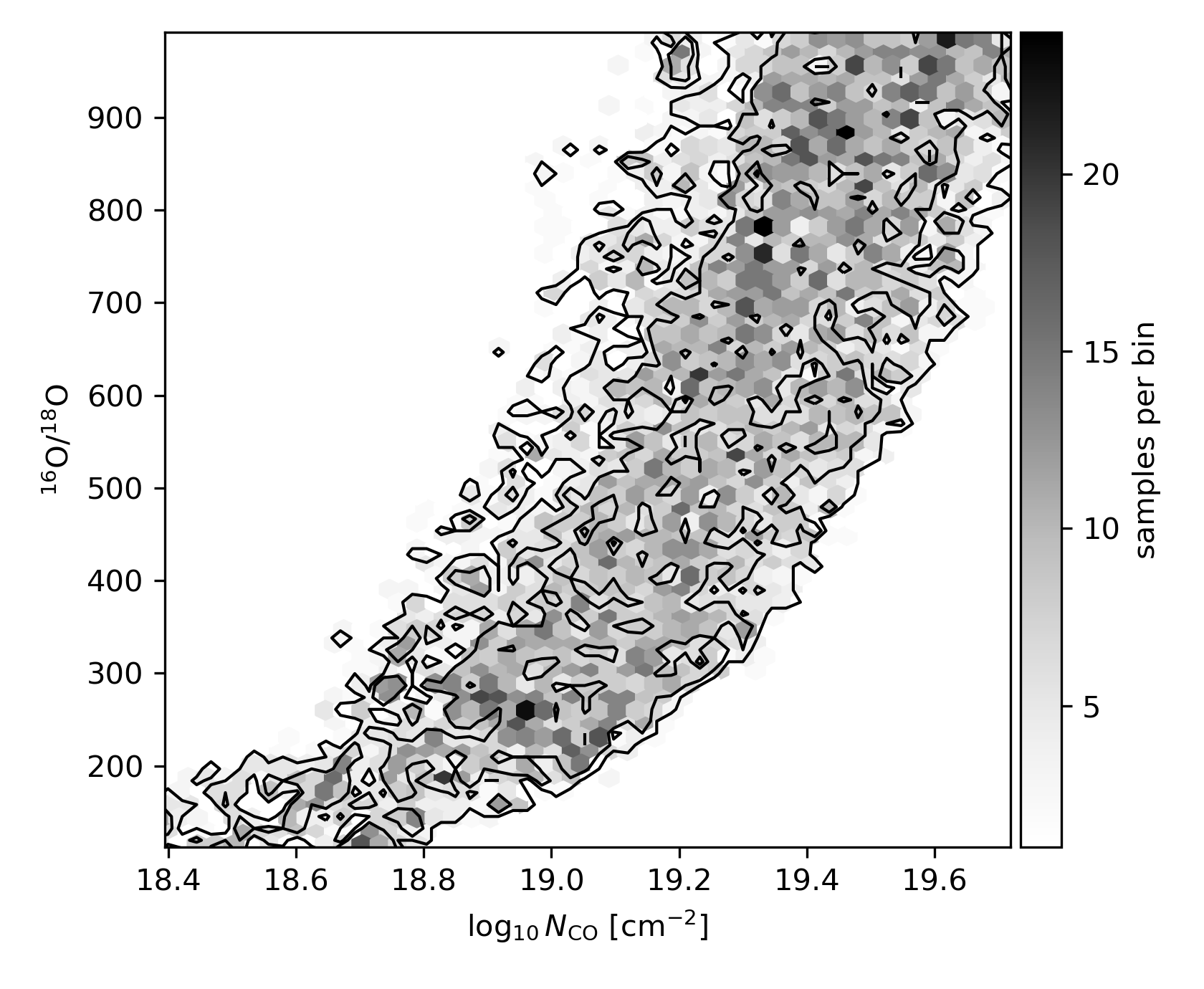}
    \caption{Two–dimensional posterior distribution of $^{16}$O/$^{18}$O versus $\log N_{\rm CO}$ from the MCMC analysis. The color scale indicates the number of samples per bin; overplotted contours mark the 68\% and 95\% credible regions. The elongated ridge illustrates the strong positive covariance between $^{16}$O/$^{18}$O and $N_{\rm CO}$: larger oxygen isotope ratios can be compensated by higher CO columns while preserving the C$^{18}$O(1--0) flux.
    }
    \label{fig:covar_R16_Nco}
\end{figure}

\paragraph{Two-dimensional covariance between $^{16}$O/$^{18}$O and $N_{\rm CO}$.}
To visualize the specific degeneracy between the oxygen isotopic ratio and the CO column density, Figure~\ref{fig:covar_R16_Nco} shows the two–dimensional posterior for $^{16}$O/$^{18}$O and $\log N_{\rm CO}$ from the same MCMC chains used in Figure~\ref{fig:mcmc_corner}. The distribution forms a ridge that runs from $\log_{10} N_{\rm CO} \sim 18.5$ with $^{16}$O/$^{18}$O $\sim 200$ up to $\log_{10} N_{\rm CO} \sim 19.6$ with $^{16}$O/$^{18}$O $\gtrsim 800$. This structure reflects the fact that, at the inferred optical depths of C$^{18}$O(1--0), an increase in $^{16}$O/$^{18}$O (which reduces $N_{\rm C^{18}O}$ at fixed $N_{\rm CO}$) can be compensated by a higher total CO column in order to match the observed C$^{18}$O intensity. This covariance explains why $^{16}$O/$^{18}$O is less tightly constrained than $^{12}$C/$^{13}$C and $^{18}$O/$^{17}$O, whose posteriors show minimal correlation with $N_{\rm CO}$.

\section{Calibration Sensitivity}
\label{app:calib}

We quantified how absolute–flux systematics propagate into the posteriors with two bracketing tests that leave the
modeling and priors unchanged and only rescale the four measured, beam–averaged integrated intensities used in the MCMC.

\paragraph{Global $\pm 10\%$ rescaling.}
All four lines were multiplied by the same gain factor ($g=0.90,\,1.00,\,1.10$). Relative to the baseline ($g=1.00$), the median shifts were small:
$\Delta(^{12}\mathrm{C}/^{13}\mathrm{C})=\{-1.281,+0.507\}$,
$\Delta(^{18}\mathrm{O}/^{17}\mathrm{O})=\{-0.001,+0.016\}$,
$\Delta(^{16}\mathrm{O}/^{18}\mathrm{O})=\{-33.410,+14.826\}$,
and $\Delta\log N_{\rm CO}=\{-0.054,+0.099\}$.
As expected, a global gain leaves the within–band, same–transition ratios that set $^{12}\mathrm{C}/^{13}\mathrm{C}$ and $^{18}\mathrm{O}/^{17}\mathrm{O}$ essentially unchanged and primarily rescales $N_{\rm CO}$.

\paragraph{Per–band $\pm 10\%$ test (worst–case inter–band mismatch).}
We scaled Band~3 (1--0 lines: C$^{18}$O, C$^{17}$O) and Band~6 (2--1 lines: $^{13}$C$^{18}$O, C$^{17}$O) in opposite
directions to bracket a realistic inter–band calibration offset:
$(g_{\rm B3},g_{\rm B6})=(1.10,0.90)$ and $(0.90,1.10)$.
Median shifts relative to the baseline were
$\Delta(^{12}\mathrm{C}/^{13}\mathrm{C})=\{+29.831,\,-16.296\}$,
$\Delta(^{18}\mathrm{O}/^{17}\mathrm{O})=\{+0.537,\,-0.455\}$,
$\Delta(^{16}\mathrm{O}/^{18}\mathrm{O})=\{-30.321,\,+29.024\}$,
and $\Delta\log N_{\rm CO}=\{-0.100,\,+0.169\}$.
These runs deliberately force a strong mismatch between the bands. The relatively large response of $^{12}\mathrm{C}/^{13}\mathrm{C}$ arises because the likelihood is defined on all four absolute fluxes, which couples the Band~3 and Band~6 data through shared parameters ($N_{\rm CO}$, $T_{\rm kin}$, $\Delta V$), even though the main observational leverage on $^{12}\mathrm{C}/^{13}\mathrm{C}$ comes from the same–transition ratio C$^{17}$O/$^{13}$C$^{18}$O(2--1) within Band~6.

\begin{table*}[t]
\centering
\caption{Calibration sensitivity: median shifts relative to baseline ($g=1.00$).}
\label{tab:cal_sens}
\begin{tabular}{lcccc}
\hline
Scenario & $\Delta(^{12}\mathrm{C}/^{13}\mathrm{C})$ & $\Delta(^{18}\mathrm{O}/^{17}\mathrm{O})$ & $\Delta(^{16}\mathrm{O}/^{18}\mathrm{O})$ & $\Delta\log N_{\rm CO}$ \\
\hline
Global $g=0.90$ & $-1.281$ & $-0.001$ & $-33.410$ & $-0.054$ \\
Global $g=1.10$ & $+0.507$ & $+0.016$ & $+14.826$ & $+0.099$ \\
Per-band $(g_{\rm B3},g_{\rm B6})=(1.10,0.90)$ & $+29.831$ & $+0.537$ & $-30.321$ & $-0.100$ \\
Per-band $(g_{\rm B3},g_{\rm B6})=(0.90,1.10)$ & $-16.296$ & $-0.455$ & $+29.024$ & $+0.169$ \\
\hline
\end{tabular}
\end{table*}

\paragraph{Summary.}
All the corresponding median shifts are summarized in Table~\ref{tab:cal_sens}. The global test leaves $^{12}\mathrm{C}/^{13}\mathrm{C}$ and $^{18}\mathrm{O}/^{17}\mathrm{O}$
stable while mainly shifting $N_{\rm CO}$. 
Even under these deliberately extreme inter–band offsets, the inferred $^{12}$C/$^{13}$C and
$^{18}$O/$^{17}$O values remain statistically consistent with both the baseline solution and the
local ISM within their quoted uncertainties, while $^{16}$O/$^{18}$O is further broadened but
already poorly constrained. Together with the global $\pm 10\%$ scaling tests, these results show
that our main isotopic conclusions, which rely on same–transition ratios in the optically thin
outer disk, are robust against 5--10\% absolute flux calibration systematics.

\bibliography{main}{}

@ARTICLE{Astropy2022,
       author = {{Astropy Collaboration} and {Price-Whelan}, Adrian M. and {Lim}, Pey Lian and {Earl}, Nicholas and {Starkman}, Nathaniel and {Bradley}, Larry and {Shupe}, David L. and {Patil}, Aarya A. and {Corrales}, Lia and {Brasseur}, C.~E. and {N{\"o}the}, Maximilian and {Donath}, Axel and {Tollerud}, Erik and {Morris}, Brett M. and {Ginsburg}, Adam and {Vaher}, Eero and {Weaver}, Benjamin A. and {Tocknell}, James and {Jamieson}, William and {van Kerkwijk}, Marten H. and {Robitaille}, Thomas P. and {Merry}, Bruce and {Bachetti}, Matteo and {G{\"u}nther}, H. Moritz and {Aldcroft}, Thomas L. and {Alvarado-Montes}, Jaime A. and {Archibald}, Anne M. and {B{\'o}di}, Attila and {Bapat}, Shreyas and {Barentsen}, Geert and {Baz{\'a}n}, Juanjo and {Biswas}, Manish and {Boquien}, M{\'e}d{\'e}ric and {Burke}, D.~J. and {Cara}, Daria and {Cara}, Mihai and {Conroy}, Kyle E. and {Conseil}, Simon and {Craig}, Matthew W. and {Cross}, Robert M. and {Cruz}, Kelle L. and {D'Eugenio}, Francesco and {Dencheva}, Nadia and {Devillepoix}, Hadrien A.~R. and {Dietrich}, J{\"o}rg P. and {Eigenbrot}, Arthur Davis and {Erben}, Thomas and {Ferreira}, Leonardo and {Foreman-Mackey}, Daniel and {Fox}, Ryan and {Freij}, Nabil and {Garg}, Suyog and {Geda}, Robel and {Glattly}, Lauren and {Gondhalekar}, Yash and {Gordon}, Karl D. and {Grant}, David and {Greenfield}, Perry and {Groener}, Austen M. and {Guest}, Steve and {Gurovich}, Sebastian and {Handberg}, Rasmus and {Hart}, Akeem and {Hatfield-Dodds}, Zac and {Homeier}, Derek and {Hosseinzadeh}, Griffin and {Jenness}, Tim and {Jones}, Craig K. and {Joseph}, Prajwel and {Kalmbach}, J. Bryce and {Karamehmetoglu}, Emir and {Ka{\l}uszy{\'n}ski}, Miko{\l}aj and {Kelley}, Michael S.~P. and {Kern}, Nicholas and {Kerzendorf}, Wolfgang E. and {Koch}, Eric W. and {Kulumani}, Shankar and {Lee}, Antony and {Ly}, Chun and {Ma}, Zhiyuan and {MacBride}, Conor and {Maljaars}, Jakob M. and {Muna}, Demitri and {Murphy}, N.~A. and {Norman}, Henrik and {O'Steen}, Richard and {Oman}, Kyle A. and {Pacifici}, Camilla and {Pascual}, Sergio and {Pascual-Granado}, J. and {Patil}, Rohit R. and {Perren}, Gabriel I. and {Pickering}, Timothy E. and {Rastogi}, Tanuj and {Roulston}, Benjamin R. and {Ryan}, Daniel F. and {Rykoff}, Eli S. and {Sabater}, Jose and {Sakurikar}, Parikshit and {Salgado}, Jes{\'u}s and {Sanghi}, Aniket and {Saunders}, Nicholas and {Savchenko}, Volodymyr and {Schwardt}, Ludwig and {Seifert-Eckert}, Michael and {Shih}, Albert Y. and {Jain}, Anany Shrey and {Shukla}, Gyanendra and {Sick}, Jonathan and {Simpson}, Chris and {Singanamalla}, Sudheesh and {Singer}, Leo P. and {Singhal}, Jaladh and {Sinha}, Manodeep and {Sip{\H{o}}cz}, Brigitta M. and {Spitler}, Lee R. and {Stansby}, David and {Streicher}, Ole and {{\v{S}}umak}, Jani and {Swinbank}, John D. and {Taranu}, Dan S. and {Tewary}, Nikita and {Tremblay}, Grant R. and {de Val-Borro}, Miguel and {Van Kooten}, Samuel J. and {Vasovi{\'c}}, Zlatan and {Verma}, Shresth and {de Miranda Cardoso}, Jos{\'e} Vin{\'\i}cius and {Williams}, Peter K.~G. and {Wilson}, Tom J. and {Winkel}, Benjamin and {Wood-Vasey}, W.~M. and {Xue}, Rui and {Yoachim}, Peter and {Zhang}, Chen and {Zonca}, Andrea and {Astropy Project Contributors}},
        title = "{The Astropy Project: Sustaining and Growing a Community-oriented Open-source Project and the Latest Major Release (v5.0) of the Core Package}",
      journal = {\apj},
         year = 2022,
        month = aug,
       volume = {935},
       number = {2},
          eid = {167},
        pages = {167},
          doi = {10.3847/1538-4357/ac7c74},
archivePrefix = {arXiv},
       eprint = {2206.14220},
 primaryClass = {astro-ph.IM},
       adsurl = {https://ui.adsabs.harvard.edu/abs/2022ApJ...935..167A}
}

@ARTICLE{Astropy2018,
       author = {{Astropy Collaboration} and {Price-Whelan}, A.~M. and {Sip{\H{o}}cz}, B.~M. and {G{\"u}nther}, H.~M. and {Lim}, P.~L. and {Crawford}, S.~M. and {Conseil}, S. and {Shupe}, D.~L. and {Craig}, M.~W. and {Dencheva}, N. and {Ginsburg}, A. and {VanderPlas}, J.~T. and {Bradley}, L.~D. and {P{\'e}rez-Su{\'a}rez}, D. and {de Val-Borro}, M. and {Aldcroft}, T.~L. and {Cruz}, K.~L. and {Robitaille}, T.~P. and {Tollerud}, E.~J. and {Ardelean}, C. and {Babej}, T. and {Bach}, Y.~P. and {Bachetti}, M. and {Bakanov}, A.~V. and {Bamford}, S.~P. and {Barentsen}, G. and {Barmby}, P. and {Baumbach}, A. and {Berry}, K.~L. and {Biscani}, F. and {Boquien}, M. and {Bostroem}, K.~A. and {Bouma}, L.~G. and {Brammer}, G.~B. and {Bray}, E.~M. and {Breytenbach}, H. and {Buddelmeijer}, H. and {Burke}, D.~J. and {Calderone}, G. and {Cano Rodr{\'\i}guez}, J.~L. and {Cara}, M. and {Cardoso}, J.~V.~M. and {Cheedella}, S. and {Copin}, Y. and {Corrales}, L. and {Crichton}, D. and {D'Avella}, D. and {Deil}, C. and {Depagne}, {\'E}. and {Dietrich}, J.~P. and {Donath}, A. and {Droettboom}, M. and {Earl}, N. and {Erben}, T. and {Fabbro}, S. and {Ferreira}, L.~A. and {Finethy}, T. and {Fox}, R.~T. and {Garrison}, L.~H. and {Gibbons}, S.~L.~J. and {Goldstein}, D.~A. and {Gommers}, R. and {Greco}, J.~P. and {Greenfield}, P. and {Groener}, A.~M. and {Grollier}, F. and {Hagen}, A. and {Hirst}, P. and {Homeier}, D. and {Horton}, A.~J. and {Hosseinzadeh}, G. and {Hu}, L. and {Hunkeler}, J.~S. and {Ivezi{\'c}}, {\v{Z}}. and {Jain}, A. and {Jenness}, T. and {Kanarek}, G. and {Kendrew}, S. and {Kern}, N.~S. and {Kerzendorf}, W.~E. and {Khvalko}, A. and {King}, J. and {Kirkby}, D. and {Kulkarni}, A.~M. and {Kumar}, A. and {Lee}, A. and {Lenz}, D. and {Littlefair}, S.~P. and {Ma}, Z. and {Macleod}, D.~M. and {Mastropietro}, M. and {McCully}, C. and {Montagnac}, S. and {Morris}, B.~M. and {Mueller}, M. and {Mumford}, S.~J. and {Muna}, D. and {Murphy}, N.~A. and {Nelson}, S. and {Nguyen}, G.~H. and {Ninan}, J.~P. and {N{\"o}the}, M. and {Ogaz}, S. and {Oh}, S. and {Parejko}, J.~K. and {Parley}, N. and {Pascual}, S. and {Patil}, R. and {Patil}, A.~A. and {Plunkett}, A.~L. and {Prochaska}, J.~X. and {Rastogi}, T. and {Reddy Janga}, V. and {Sabater}, J. and {Sakurikar}, P. and {Seifert}, M. and {Sherbert}, L.~E. and {Sherwood-Taylor}, H. and {Shih}, A.~Y. and {Sick}, J. and {Silbiger}, M.~T. and {Singanamalla}, S. and {Singer}, L.~P. and {Sladen}, P.~H. and {Sooley}, K.~A. and {Sornarajah}, S. and {Streicher}, O. and {Teuben}, P. and {Thomas}, S.~W. and {Tremblay}, G.~R. and {Turner}, J.~E.~H. and {Terr{\'o}n}, V. and {van Kerkwijk}, M.~H. and {de la Vega}, A. and {Watkins}, L.~L. and {Weaver}, B.~A. and {Whitmore}, J.~B. and {Woillez}, J. and {Zabalza}, V. and {Astropy Contributors}},
        title = "{The Astropy Project: Building an Open-science Project and Status of the v2.0 Core Package}",
      journal = {\aj},
         year = 2018,
        month = sep,
       volume = {156},
       number = {3},
          eid = {123},
        pages = {123},
          doi = {10.3847/1538-3881/aabc4f},
archivePrefix = {arXiv},
       eprint = {1801.02634},
 primaryClass = {astro-ph.IM},
       adsurl = {https://ui.adsabs.harvard.edu/abs/2018AJ....156..123A}
}

@ARTICLE{Astropy2013,
       author = {{Astropy Collaboration} and {Robitaille}, Thomas P. and
         {Tollerud}, Erik J. and {Greenfield}, Perry and {Droettboom}, Michael and
         {Bray}, Erik and {Aldcroft}, Tom and {Davis}, Matt and
         {Ginsburg}, Adam and {Price-Whelan}, Adrian M. and
         {Kerzendorf}, Wolfgang E. and {Conley}, Alexander and {Crighton}, Neil and
         {Barbary}, Kyle and {Muna}, Demitri and {Ferguson}, Henry and
         {Grollier}, Fr{\'e}d{\'e}ric and {Parikh}, Madhura M. and
         {Nair}, Prasanth H. and {Unther}, Hans M. and {Deil}, Christoph and
         {Woillez}, Julien and {Conseil}, Simon and {Kramer}, Roban and
         {Turner}, James E.~H. and {Singer}, Leo and {Fox}, Ryan and
         {Weaver}, Benjamin A. and {Zabalza}, Victor and {Edwards}, Zachary I. and
         {Azalee Bostroem}, K. and {Burke}, D.~J. and {Casey}, Andrew R. and
         {Crawford}, Steven M. and {Dencheva}, Nadia and {Ely}, Justin and
         {Jenness}, Tim and {Labrie}, Kathleen and {Lim}, Pey Lian and
         {Pierfederici}, Francesco and {Pontzen}, Andrew and {Ptak}, Andy and
         {Refsdal}, Brian and {Servillat}, Mathieu and {Streicher}, Ole},
        title = "{Astropy: A community Python package for astronomy}",
      journal = {\aap},
         year = "2013",
        month = "Oct",
       volume = {558},
          eid = {A33},
        pages = {A33},
          doi = {10.1051/0004-6361/201322068},
archivePrefix = {arXiv},
       eprint = {1307.6212},
 primaryClass = {astro-ph.IM},
       adsurl = {https://ui.adsabs.harvard.edu/abs/2013A&A...558A..33A}
}

@ARTICLE{Altwegg2019,
       author = {{Altwegg}, Kathrin and {Balsiger}, Hans and {Fuselier}, Stephen A.},
        title = "{Cometary Chemistry and the Origin of Icy Solar System Bodies: The View After Rosetta}",
      journal = {\araa},
         year = 2019,
        month = aug,
       volume = {57},
        pages = {113-155},
          doi = {10.1146/annurev-astro-091918-104409},
archivePrefix = {arXiv},
       eprint = {1908.04046},
 primaryClass = {astro-ph.EP},
       adsurl = {https://ui.adsabs.harvard.edu/abs/2019ARA&A..57..113A}
}

@ARTICLE{CASA+2022,
       author = {{CASA Team} and {Bean}, Ben and {Bhatnagar}, Sanjay and {Castro}, Sandra and {Donovan Meyer}, Jennifer and {Emonts}, Bjorn and {Garcia}, Enrique and {Garwood}, Robert and {Golap}, Kumar and {Gonzalez Villalba}, Justo and {Harris}, Pamela and {Hayashi}, Yohei and {Hoskins}, Josh and {Hsieh}, Mingyu and {Jagannathan}, Preshanth and {Kawasaki}, Wataru and {Keimpema}, Aard and {Kettenis}, Mark and {Lopez}, Jorge and {Marvil}, Joshua and {Masters}, Joseph and {McNichols}, Andrew and {Mehringer}, David and {Miel}, Renaud and {Moellenbrock}, George and {Montesino}, Federico and {Nakazato}, Takeshi and {Ott}, Juergen and {Petry}, Dirk and {Pokorny}, Martin and {Raba}, Ryan and {Rau}, Urvashi and {Schiebel}, Darrell and {Schweighart}, Neal and {Sekhar}, Srikrishna and {Shimada}, Kazuhiko and {Small}, Des and {Steeb}, Jan-Willem and {Sugimoto}, Kanako and {Suoranta}, Ville and {Tsutsumi}, Takahiro and {van Bemmel}, Ilse M. and {Verkouter}, Marjolein and {Wells}, Akeem and {Xiong}, Wei and {Szomoru}, Arpad and {Griffith}, Morgan and {Glendenning}, Brian and {Kern}, Jeff},
        title = "{CASA, the Common Astronomy Software Applications for Radio Astronomy}",
      journal = {\pasp},
         year = 2022,
        month = nov,
       volume = {134},
       number = {1041},
          eid = {114501},
        pages = {114501},
          doi = {10.1088/1538-3873/ac9642},
archivePrefix = {arXiv},
       eprint = {2210.02276},
 primaryClass = {astro-ph.IM},
       adsurl = {https://ui.adsabs.harvard.edu/abs/2022PASP..134k4501C}
}

@ARTICLE{Czekala2021,
       author = {{Czekala}, Ian and {Loomis}, Ryan A. and {Teague}, Richard and {Booth}, Alice S. and {Huang}, Jane and {Cataldi}, Gianni and {Ilee}, John D. and {Law}, Charles J. and {Walsh}, Catherine and {Bosman}, Arthur D. and {Guzm{\'a}n}, Viviana V. and {Le Gal}, Romane and {{\"O}berg}, Karin I. and {Yamato}, Yoshihide and {Aikawa}, Yuri and {Andrews}, Sean M. and {Bae}, Jaehan and {Bergin}, Edwin A. and {Bergner}, Jennifer B. and {Cleeves}, L. Ilsedore and {Kurtovic}, Nicolas T. and {M{\'e}nard}, Fran{\c{c}}ois and {Nomura}, Hideko and {P{\'e}rez}, Laura M. and {Qi}, Chunhua and {Schwarz}, Kamber R. and {Tsukagoshi}, Takashi and {Waggoner}, Abygail R. and {Wilner}, David J. and {Zhang}, Ke},
        title = "{Molecules with ALMA at Planet-forming Scales (MAPS). II. CLEAN Strategies for Synthesizing Images of Molecular Line Emission in Protoplanetary Disks}",
      journal = {\apjs},
         year = 2021,
        month = nov,
       volume = {257},
       number = {1},
          eid = {2},
        pages = {2},
          doi = {10.3847/1538-4365/ac1430},
archivePrefix = {arXiv},
       eprint = {2109.06188},
 primaryClass = {astro-ph.EP},
       adsurl = {https://ui.adsabs.harvard.edu/abs/2021ApJS..257....2C}
}

@ARTICLE{Deng2023,
       author = {{Deng}, Dingshan and {Ruaud}, Maxime and {Gorti}, Uma and {Pascucci}, Ilaria},
        title = "{DiskMINT: A Tool to Estimate Disk Masses with CO Isotopologues}",
      journal = {\apj},
         year = 2023,
        month = sep,
       volume = {954},
       number = {2},
          eid = {165},
        pages = {165},
          doi = {10.3847/1538-4357/acdfcc},
archivePrefix = {arXiv},
       eprint = {2307.02657},
 primaryClass = {astro-ph.EP},
       adsurl = {https://ui.adsabs.harvard.edu/abs/2023ApJ...954..165D}
}

@ARTICLE{Deng2025,
       author = {{Deng}, Dingshan and {Gorti}, Uma and {Pascucci}, Ilaria and {Ruaud}, Maxime},
        title = "{DiskMINT: Self-Consistent Thermochemical Disk Models with Radially Varying Gas and Dust -- Application to the Massive, CO-Rich Disk of IM Lup}",
      journal = {arXiv e-prints},
         year = 2025,
        month = sep,
          eid = {arXiv:2509.15487},
        pages = {arXiv:2509.15487},
          doi = {10.48550/arXiv.2509.15487},
archivePrefix = {arXiv},
       eprint = {2509.15487},
 primaryClass = {astro-ph.EP},
       adsurl = {https://ui.adsabs.harvard.edu/abs/2025arXiv250915487D}
}

@ARTICLE{emcee,
       author = {{Foreman-Mackey}, Daniel and {Hogg}, David W. and {Lang}, Dustin and
        {Goodman}, Jonathan},
        title = "{emcee: The MCMC Hammer}",
      journal = {Publications of the Astronomical Society of the Pacific},
         year = 2013,
        month = Mar,
       volume = {125},
        pages = {306},
          doi = {10.1086/670067},
archivePrefix = {arXiv},
       eprint = {1202.3665},
       adsurl = {https://ui.adsabs.harvard.edu/#abs/2013PASP..125..306F}
}

@ARTICLE{Furuya2022,
       author = {{Furuya}, Kenji and {Tsukagoshi}, Takashi and {Qi}, Chunhua and {Nomura}, Hideko and {Cleeves}, L. Ilsedore and {Lee}, Seokho and {Yoshida}, Tomohiro C.},
        title = "{Detection of HC$^{18}$O$^{+}$ in a Protoplanetary Disk: Exploring Oxygen Isotope Fractionation of CO}",
      journal = {\apj},
         year = 2022,
        month = feb,
       volume = {926},
       number = {2},
          eid = {148},
        pages = {148},
          doi = {10.3847/1538-4357/ac45ff},
archivePrefix = {arXiv},
       eprint = {2201.00935},
 primaryClass = {astro-ph.EP},
       adsurl = {https://ui.adsabs.harvard.edu/abs/2022ApJ...926..148F}
}

@ARTICLE{Jorsater1995,
       author = {{Jorsater}, Steven and {van Moorsel}, Gustaaf A.},
        title = "{High Resolution Neutral Hydrogen Observations of the Barred Spiral Galaxy NGC 1365}",
      journal = {\aj},
         year = 1995,
        month = nov,
       volume = {110},
        pages = {2037},
          doi = {10.1086/117668},
       adsurl = {https://ui.adsabs.harvard.edu/abs/1995AJ....110.2037J}
}

@ARTICLE{Law2021,
       author = {{Law}, Charles J. and {Teague}, Richard and {Loomis}, Ryan A. and {Bae}, Jaehan and {{\"O}berg}, Karin I. and {Czekala}, Ian and {Andrews}, Sean M. and {Aikawa}, Yuri and {Alarc{\'o}n}, Felipe and {Bergin}, Edwin A. and {Bergner}, Jennifer B. and {Booth}, Alice S. and {Bosman}, Arthur D. and {Calahan}, Jenny K. and {Cataldi}, Gianni and {Cleeves}, L. Ilsedore and {Furuya}, Kenji and {Guzm{\'a}n}, Viviana V. and {Huang}, Jane and {Ilee}, John D. and {Le Gal}, Romane and {Liu}, Yao and {Long}, Feng and {M{\'e}nard}, Fran{\c{c}}ois and {Nomura}, Hideko and {P{\'e}rez}, Laura M. and {Qi}, Chunhua and {Schwarz}, Kamber R. and {Soto}, Daniela and {Tsukagoshi}, Takashi and {Yamato}, Yoshihide and {van't Hoff}, Merel L.~R. and {Walsh}, Catherine and {Wilner}, David J. and {Zhang}, Ke},
        title = "{Molecules with ALMA at Planet-forming Scales (MAPS). IV. Emission Surfaces and Vertical Distribution of Molecules}",
      journal = {\apjs},
         year = 2021,
        month = nov,
       volume = {257},
       number = {1},
          eid = {4},
        pages = {4},
          doi = {10.3847/1538-4365/ac1439},
archivePrefix = {arXiv},
       eprint = {2109.06217},
 primaryClass = {astro-ph.GA},
       adsurl = {https://ui.adsabs.harvard.edu/abs/2021ApJS..257....4L}
}

@ARTICLE{Lee2024,
       author = {{Lee}, Seokho and {Nomura}, Hideko and {Furuya}, Kenji},
        title = "{Carbon Isotope Chemistry in Protoplanetary Disks: Effects of C/O Ratios}",
      journal = {\apj},
         year = 2024,
        month = jul,
       volume = {969},
       number = {1},
          eid = {41},
        pages = {41},
          doi = {10.3847/1538-4357/ad39e3},
archivePrefix = {arXiv},
       eprint = {2404.01635},
 primaryClass = {astro-ph.GA},
       adsurl = {https://ui.adsabs.harvard.edu/abs/2024ApJ...969...41L}
}

@ARTICLE{Manfroid2009,
       author = {{Manfroid}, J. and {Jehin}, E. and {Hutsem{\'e}kers}, D. and {Cochran}, A. and {Zucconi}, J. -M. and {Arpigny}, C. and {Schulz}, R. and {St{\"u}we}, J.~A. and {Ilyin}, I.},
        title = "{The CN isotopic ratios in comets}",
      journal = {\aap},
         year = 2009,
        month = aug,
       volume = {503},
       number = {2},
        pages = {613-624},
          doi = {10.1051/0004-6361/200911859},
archivePrefix = {arXiv},
       eprint = {0907.0311},
 primaryClass = {astro-ph.EP},
       adsurl = {https://ui.adsabs.harvard.edu/abs/2009A&A...503..613M}
}

@ARTICLE{Milam2005,
       author = {{Milam}, S.~N. and {Savage}, C. and {Brewster}, M.~A. and {Ziurys}, L.~M. and {Wyckoff}, S.},
        title = "{The $^{12}$C/$^{13}$C Isotope Gradient Derived from Millimeter Transitions of CN: The Case for Galactic Chemical Evolution}",
      journal = {\apj},
         year = 2005,
        month = dec,
       volume = {634},
       number = {2},
        pages = {1126-1132},
          doi = {10.1086/497123},
       adsurl = {https://ui.adsabs.harvard.edu/abs/2005ApJ...634.1126M}
}

@ARTICLE{Miotello2016,
       author = {{Miotello}, A. and {van Dishoeck}, E.~F. and {Kama}, M. and {Bruderer}, S.},
        title = "{Determining protoplanetary disk gas masses from CO isotopologues line observations}",
      journal = {\aap},
         year = 2016,
        month = oct,
       volume = {594},
          eid = {A85},
        pages = {A85},
          doi = {10.1051/0004-6361/201628159},
archivePrefix = {arXiv},
       eprint = {1605.07780},
 primaryClass = {astro-ph.SR},
       adsurl = {https://ui.adsabs.harvard.edu/abs/2016A&A...594A..85M}
}

@ARTICLE{Miotello2021,
       author = {{Miotello}, A. and {Rosotti}, G. and {Ansdell}, M. and {Facchini}, S. and {Manara}, C.~F. and {Williams}, J.~P. and {Bruderer}, S.},
        title = "{Compact disks. An explanation to faint CO emission in Lupus disks}",
      journal = {\aap},
         year = 2021,
        month = jul,
       volume = {651},
          eid = {A48},
        pages = {A48},
          doi = {10.1051/0004-6361/202140550},
archivePrefix = {arXiv},
       eprint = {2104.09109},
 primaryClass = {astro-ph.SR},
       adsurl = {https://ui.adsabs.harvard.edu/abs/2021A&A...651A..48M}
}

@INPROCEEDINGS{McMullin2007,
       author = {{McMullin}, J.~P. and {Waters}, B. and {Schiebel}, D. and {Young}, W. and {Golap}, K.},
        title = "{CASA Architecture and Applications}",
    booktitle = {Astronomical Data Analysis Software and Systems XVI},
         year = 2007,
       editor = {{Shaw}, R.~A. and {Hill}, F. and {Bell}, D.~J.},
       series = {Astronomical Society of the Pacific Conference Series},
       volume = {376},
        month = oct,
        pages = {127},
       adsurl = {https://ui.adsabs.harvard.edu/abs/2007ASPC..376..127M}
}

@ARTICLE{Qi2011,
       author = {{Qi}, Chunhua and {D'Alessio}, Paola and {{\"O}berg}, Karin I. and {Wilner}, David J. and {Hughes}, A. Meredith and {Andrews}, Sean M. and {Ayala}, Sandra},
        title = "{Resolving the CO Snow Line in the Disk around HD 163296}",
      journal = {\apj},
         year = 2011,
        month = oct,
       volume = {740},
       number = {2},
          eid = {84},
        pages = {84},
          doi = {10.1088/0004-637X/740/2/84},
archivePrefix = {arXiv},
       eprint = {1107.5061},
 primaryClass = {astro-ph.SR},
       adsurl = {https://ui.adsabs.harvard.edu/abs/2011ApJ...740...84Q}
}

@ARTICLE{Qi2024,
       author = {{Qi}, Chunhua and {Wilner}, David J.},
        title = "{Evidence for a Sharp CO Snow Line Transition in a Protoplanetary Disk and Implications for Millimeter-wave Observations of CO Isotopologues}",
      journal = {\apj},
         year = 2024,
        month = dec,
       volume = {977},
       number = {1},
          eid = {60},
        pages = {60},
          doi = {10.3847/1538-4357/ad8d55},
archivePrefix = {arXiv},
       eprint = {2410.23036},
 primaryClass = {astro-ph.EP},
       adsurl = {https://ui.adsabs.harvard.edu/abs/2024ApJ...977...60Q}
}

@ARTICLE{Hily-Blant2019,
       author = {{Hily-Blant}, P. and {Magalhaes de Souza}, V. and {Kastner}, J. and {Forveille}, T.},
        title = "{Multiple nitrogen reservoirs in a protoplanetary disk at the epoch of comet and giant planet formation}",
      journal = {\aap},
         year = 2019,
        month = dec,
       volume = {632},
          eid = {L12},
        pages = {L12},
          doi = {10.1051/0004-6361/201936750},
archivePrefix = {arXiv},
       eprint = {1911.06676},
 primaryClass = {astro-ph.GA},
       adsurl = {https://ui.adsabs.harvard.edu/abs/2019A&A...632L..12H}
}

@ARTICLE{Holdship2021,
       author = {{Holdship}, J. and {Viti}, S. and {Mart{\'\i}n}, S. and {Harada}, N. and {Mangum}, J. and {Sakamoto}, K. and {Muller}, S. and {Tanaka}, K. and {Yoshimura}, Y. and {Nakanishi}, K. and {Herrero-Illana}, R. and {M{\"u}hle}, S. and {Aladro}, R. and {Colzi}, L. and {Emig}, K.~L. and {Garc{\'\i}a-Burillo}, S. and {Henkel}, C. and {Humire}, P. and {Meier}, D.~S. and {Rivilla}, V.~M. and {van der Werf}, P.},
        title = "{The distribution and origin of C$_{2}$H in NGC 253 from ALCHEMI}",
      journal = {\aap},
         year = 2021,
        month = oct,
       volume = {654},
          eid = {A55},
        pages = {A55},
          doi = {10.1051/0004-6361/202141233},
archivePrefix = {arXiv},
       eprint = {2107.04580},
 primaryClass = {astro-ph.GA},
       adsurl = {https://ui.adsabs.harvard.edu/abs/2021A&A...654A..55H}
}

@ARTICLE{Hunter2007,
       author = {{Hunter}, John D.},
        title = "{Matplotlib: A 2D Graphics Environment}",
      journal = {Computing in Science and Engineering},
         year = 2007,
        month = may,
       volume = {9},
       number = {3},
        pages = {90-95},
          doi = {10.1109/MCSE.2007.55},
       adsurl = {https://ui.adsabs.harvard.edu/abs/2007CSE.....9...90H}
}

@ARTICLE{Rampinelli2025,
       author = {{Rampinelli}, L. and {Facchini}, S. and {Leemker}, M. and {Curone}, P. and {Benisty}, M. and {{\"O}berg}, K.~I. and {Teague}, R. and {Andrews}, S. and {Bae}, J. and {Law}, C.~J. and {Portilla-Revelo}, B.},
        title = "{Radial variations in the nitrogen, carbon, and hydrogen fractionation in the PDS 70 planet-hosting disk}",
      journal = {\aap},
         year = 2025,
        month = jun,
       volume = {698},
          eid = {A115},
        pages = {A115},
          doi = {10.1051/0004-6361/202554172},
archivePrefix = {arXiv},
       eprint = {2504.03833},
 primaryClass = {astro-ph.EP},
       adsurl = {https://ui.adsabs.harvard.edu/abs/2025A&A...698A.115R}
}

@ARTICLE{Ruaud2022,
       author = {{Ruaud}, Maxime and {Gorti}, Uma and {Hollenbach}, David J.},
        title = "{C$^{18}$O Emission as an Effective Measure of Gas Masses of Protoplanetary Disks}",
      journal = {\apj},
         year = 2022,
        month = jan,
       volume = {925},
       number = {1},
          eid = {49},
        pages = {49},
          doi = {10.3847/1538-4357/ac3826},
archivePrefix = {arXiv},
       eprint = {2111.05833},
 primaryClass = {astro-ph.GA},
       adsurl = {https://ui.adsabs.harvard.edu/abs/2022ApJ...925...49R}
}

@ARTICLE{Rubin2019,
       author = {{Rubin}, Martin and {Altwegg}, Kathrin and {Balsiger}, Hans and {Berthelier}, Jean-Jacques and {Combi}, Michael R. and {De Keyser}, Johan and {Drozdovskaya}, Maria and {Fiethe}, Bj{\"o}rn and {Fuselier}, Stephen A. and {Gasc}, S{\'e}bastien and {Gombosi}, Tamas I. and {H{\"a}nni}, Nora and {Hansen}, Kenneth C. and {Mall}, Urs and {R{\`e}me}, Henri and {Schroeder}, Isaac R.~H.~G. and {Schuhmann}, Markus and {S{\'e}mon}, Thierry and {Waite}, Jack H. and {Wampfler}, Susanne F. and {Wurz}, Peter},
        title = "{Elemental and molecular abundances in comet 67P/Churyumov-Gerasimenko}",
      journal = {\mnras},
         year = 2019,
        month = oct,
       volume = {489},
       number = {1},
        pages = {594-607},
          doi = {10.1093/mnras/stz2086},
archivePrefix = {arXiv},
       eprint = {1907.11044},
 primaryClass = {astro-ph.EP},
       adsurl = {https://ui.adsabs.harvard.edu/abs/2019MNRAS.489..594R}
}

@ARTICLE{Schoier2005,
       author = {{Sch{\"o}ier}, F.~L. and {van der Tak}, F.~F.~S. and {van Dishoeck}, E.~F. and {Black}, J.~H.},
        title = "{An atomic and molecular database for analysis of submillimetre line observations}",
      journal = {\aap},
         year = 2005,
        month = mar,
       volume = {432},
       number = {1},
        pages = {369-379},
          doi = {10.1051/0004-6361:20041729},
archivePrefix = {arXiv},
       eprint = {astro-ph/0411110},
 primaryClass = {astro-ph},
       adsurl = {https://ui.adsabs.harvard.edu/abs/2005A&A...432..369S}
}

@software{Teague2018,
       author = {{Teague}, Richard and {Foreman-Mackey}, Daniel},
        title = "{bettermoments: A robust method to measure line centroids}",
         year = 2018,
        month = sep,
          eid = {10.5281/zenodo.1419754},
          doi = {10.5281/zenodo.1419754},
      version = {v1.0},
    publisher = {Zenodo},
       adsurl = {https://ui.adsabs.harvard.edu/abs/2018zndo...1419754T}
}

@ARTICLE{Teague2019,
       author = {{Teague}, Richard},
        title = "{GoFish: Fishing for Line Observations in Protoplanetary Disks}",
      journal = {The Journal of Open Source Software},
         year = 2019,
        month = sep,
       volume = {4},
       number = {41},
          eid = {1632},
        pages = {1632},
          doi = {10.21105/joss.01632},
       adsurl = {https://ui.adsabs.harvard.edu/abs/2019JOSS....4.1632T}
}

@ARTICLE{vanderTak2007,
       author = {{van der Tak}, F.~F.~S. and {Black}, J.~H. and {Sch{\"o}ier}, F.~L. and {Jansen}, D.~J. and {van Dishoeck}, E.~F.},
        title = "{A computer program for fast non-LTE analysis of interstellar line spectra. With diagnostic plots to interpret observed line intensity ratios}",
      journal = {\aap},
         year = 2007,
        month = jun,
       volume = {468},
       number = {2},
        pages = {627-635},
          doi = {10.1051/0004-6361:20066820},
archivePrefix = {arXiv},
       eprint = {0704.0155},
 primaryClass = {astro-ph},
       adsurl = {https://ui.adsabs.harvard.edu/abs/2007A&A...468..627V}
}

@ARTICLE{vanderWalt2011,
       author = {{van der Walt}, St{\'e}fan and {Colbert}, S. Chris and {Varoquaux}, Ga{\"e}l},
        title = "{The NumPy Array: A Structure for Efficient Numerical Computation}",
      journal = {Computing in Science and Engineering},
         year = 2011,
        month = mar,
       volume = {13},
       number = {2},
        pages = {22-30},
          doi = {10.1109/MCSE.2011.37},
archivePrefix = {arXiv},
       eprint = {1102.1523},
 primaryClass = {cs.MS},
       adsurl = {https://ui.adsabs.harvard.edu/abs/2011CSE....13b..22V}
}

@ARTICLE{Visser2009,
       author = {{Visser}, R. and {van Dishoeck}, E.~F. and {Black}, J.~H.},
        title = "{The photodissociation and chemistry of CO isotopologues: applications to interstellar clouds and circumstellar disks}",
      journal = {\aap},
         year = 2009,
        month = aug,
       volume = {503},
       number = {2},
        pages = {323-343},
          doi = {10.1051/0004-6361/200912129},
archivePrefix = {arXiv},
       eprint = {0906.3699},
 primaryClass = {astro-ph.GA},
       adsurl = {https://ui.adsabs.harvard.edu/abs/2009A&A...503..323V}
}

@ARTICLE{Wilson1999,
       author = {{Wilson}, T.~L.},
        title = "{Isotopes in the interstellar medium and circumstellar envelopes}",
      journal = {Reports on Progress in Physics},
         year = 1999,
        month = feb,
       volume = {62},
       number = {2},
        pages = {143-185},
          doi = {10.1088/0034-4885/62/2/002},
       adsurl = {https://ui.adsabs.harvard.edu/abs/1999RPPh...62..143W}
}

@ARTICLE{Yang2010,
       author = {{Yang}, Benhui and {Stancil}, P.~C. and {Balakrishnan}, N. and {Forrey}, R.~C.},
        title = "{Rotational Quenching of CO due to H$_{2}$ Collisions}",
      journal = {\apj},
         year = 2010,
        month = aug,
       volume = {718},
       number = {2},
        pages = {1062-1069},
          doi = {10.1088/0004-637X/718/2/1062},
archivePrefix = {arXiv},
       eprint = {1004.3923},
 primaryClass = {astro-ph.SR},
       adsurl = {https://ui.adsabs.harvard.edu/abs/2010ApJ...718.1062Y}
}

@ARTICLE{Yoshida2022,
       author = {{Yoshida}, Tomohiro C. and {Nomura}, Hideko and {Furuya}, Kenji and {Tsukagoshi}, Takashi and {Lee}, Seokho},
        title = "{A New Method for Direct Measurement of Isotopologue Ratios in Protoplanetary Disks: A Case Study of the $^{12}$CO/$^{13}$CO Ratio in the TW Hya Disk}",
      journal = {\apj},
         year = 2022,
        month = jun,
       volume = {932},
       number = {2},
          eid = {126},
        pages = {126},
          doi = {10.3847/1538-4357/ac6efb},
archivePrefix = {arXiv},
       eprint = {2204.08330},
 primaryClass = {astro-ph.EP},
       adsurl = {https://ui.adsabs.harvard.edu/abs/2022ApJ...932..126Y}
}

@ARTICLE{Yoshida2024,
       author = {{Yoshida}, Tomohiro C. and {Nomura}, Hideko and {Furuya}, Kenji and {Teague}, Richard and {Law}, Charles J. and {Tsukagoshi}, Takashi and {Lee}, Seokho and {Rab}, Christian and {{\"O}berg}, Karin I. and {Loomis}, Ryan A.},
        title = "{The First Spatially Resolved Detection of $^{13}$CN in a Protoplanetary Disk and Evidence for Complex Carbon Isotope Fractionation}",
      journal = {\apj},
         year = 2024,
        month = may,
       volume = {966},
       number = {1},
          eid = {63},
        pages = {63},
          doi = {10.3847/1538-4357/ad2fb4},
archivePrefix = {arXiv},
       eprint = {2403.00626},
 primaryClass = {astro-ph.EP},
       adsurl = {https://ui.adsabs.harvard.edu/abs/2024ApJ...966...63Y}
}
\bibliographystyle{aasjournalv7}



\end{document}